\documentstyle[epsf,12pt]{article}

\catcode`@=11
\def\un#1{\relax\ifmmode\@@underline#1\else
        $\@@underline{\hbox{#1}}$\relax\fi}
\catcode`@=12

\def\a{\alpha}
\def\b{\beta}
\def\c{\chi}
\def\d{\delta}

\def\f{\phi}
\def\g{\gamma}
\def\h{\eta}

\def\j{\psi}

\def\l{\lambda}
\def\m{\mu}
\def\n{\nu}
\def\o{\omega}
\def\p{\pi}
\def\q{\theta}
\def\r{\rho}

\def\t{\tau}

\def\z{\zeta}

\def\F{\Phi}
\def\G{\Gamma}
\def\J{\Psi}
\def\L{\Lambda}
\def\O{\Omega}

\def\S{\Sigma}

\def\ve{\varepsilon}

\def\cg{{\cal G}}
\def\ch{{\cal H}}

\def\ck{{\cal K}}

\def\cm{{\cal M}}
\def\cn{{\cal N}}
\def\co{{\cal O}}

\def\bo{{\hbox{\large$\Box$}}}                 % D'Alembertian
\def\Bo{\raise -1pt\hbox{\bo{\hskip 0.03in}}}  % improved D'Alembertian

\def\pa{\partial}                                       % curly d
\def\de{\nabla}                                         % del
\def\TH{{\raise.2ex\hbox{$\displaystyle \bigodot$}\mskip-4.7mu \llap H \;}}
\def\face{{\raise.2ex\hbox{$\displaystyle \bigodot$}\mskip-2.2mu \llap {$\ddot
        \smile$}}}                                      % happy face
\def\dg{\sp\dagger}                                     % hermitian conjugate
\def\sp#1{{}^{#1}}                              % superscript (unaligned)
\def\Hat#1{\widehat{#1}}                        % big hat
\def\Bar#1{\overline{#1}}                       % big bar
\def\ket#1{\left| #1\right\rangle}              % | >
\def\leftrightarrowfill{$\mathsurround=0pt \mathord\leftarrow \mkern-6mu
        \cleaders\hbox{$\mkern-2mu \mathord- \mkern-2mu$}\hfill
        \mkern-6mu \mathord\rightarrow$}
\def\dvec#1{\vbox{\ialign{##\crcr
        \leftrightarrowfill\crcr\noalign{\kern-1pt\nointerlineskip}
        $\hfil\displaystyle{#1}\hfil$\crcr}}}           % <--> accent
\def\frac#1#2{{\textstyle{#1\over\vphantom2\smash{\raise.20ex
        \hbox{$\scriptstyle{#2}$}}}}}                   % fraction
\def\sfrac#1#2{{\vphantom1\smash{\lower.5ex\hbox{\small$#1$}}\over
        \vphantom1\smash{\raise.4ex\hbox{\small$#2$}}}} % alternate fraction
\def\bfrac#1#2{{\vphantom1\smash{\lower.5ex\hbox{$#1$}}\over
        \vphantom1\smash{\raise.3ex\hbox{$#2$}}}}       % "
\def\afrac#1#2{{\vphantom1\smash{\lower.5ex\hbox{$#1$}}\over#2}}    % "
\def\[{\lfloor{\hskip 0.35pt}\!\!\!\lceil}  % improved commutator left bracket
\def\]{\rfloor{\hskip 0.35pt}\!\!\!\rceil} % improved commutator right bracket
\def\ud#1#2{^{#1}{}_{#2}}
\def\dud#1#2#3{_{#1}{}^{#2}{}_{#3}}

\def\fracm#1#2{\hbox{\large{${\frac{{#1}}{{#2}}}$}}}
\def\fracmm#1#2{{{#1}\over{#2}}}                        % fractions
\def\half{{\fracm12}}
\def\ha{\half}
\def\fracmm#1#2{{{#1}\over{#2}}}

\def\tr{{\rm tr}}                                    % traces

\def\un{{\underline n}}

\def\low#1{{\raise -3pt\hbox{${\hskip 0.75pt}\!_{#1}$}}}  % big indices
\def\Low#1{{\raise -8pt\hbox{$\!\!\!\!\!\!_{#1}$}}~\;}
\def\low#1{{\raise -3pt\hbox{${\hskip 1.0pt}\!_{#1}$}}}

\def\Hat#1{\widehat{#1}}

\def\plpl{{+\!\!\!\!\!{\hskip 0.009in}{\raise -1.0pt\hbox{$_+$}}
{\hskip 0.0008in}}}

\def\mimi{{-\!\!\!\!\!{\hskip 0.009in}{\raise -1.0pt\hbox{$_-$}}
{\hskip 0.0008in}}}

\def\-{{\hskip 1.5pt}\hbox{-}}

\def\ip{{=\!\!\! \mid}}

\newskip\humongous \humongous=0pt plus 1000pt minus 1000pt
\def\caja{\mathsurround=0pt}
\def\eqalign#1{\,\vcenter{\openup2\jot \caja
        \ialign{\strut \hfil$\displaystyle{##}$&$
        \displaystyle{{}##}$\hfil\crcr#1\crcr}}\,}
\newif\ifdtup

\def\NPB{{\sf Nucl. Phys. }{\bf B}}
\def\PL{{\sf Phys. Lett. }}

\def\PRD{{\sf Phys. Rev. }{\bf D}}
\def\CQG{{\sf Class. Quantum Grav. }}

\def\CMP{{\sf Commun. Math. Phys. }}

\def\ref#1{$\sp{#1)}$}

\parskip=\medskipamount           % space between paragraphs (LaTeX)
\thicklines                     % thick straight lines for pictures (LaTeX)

\begin{document}

% =========================== title page ==================================

\thispagestyle{empty}             % no heading or foot on title page (LaTeX)

\def\border{                                            % border
        \setlength{\unitlength}{1mm}
        \newcount\xco
        \newcount\yco
        \xco=-24
        \yco=12
        \begin{picture}(140,0)
        \put(-20,11){\tiny Institut f\"ur Theoretische Physik Universit\"at
Hannover~~ Institut f\"ur Theoretische Physik Universit\"at Hannover~~
Institut f\"ur Theoretische Physik Hannover}
        \put(-20,-241.5){\tiny Department of Physics University of Maryland~
Department of Physics University of Maryland~ Department of Physics
University of Maryland~ Department of Physics}
        \end{picture}
        \par\vskip-8mm}

\def\headpic{                                           % UH heading
        \indent
        \setlength{\unitlength}{.8mm}
        \thinlines
        \par
        \begin{picture}(29,16)
        \put(75,16){\line(1,0){4}}
        \put(80,16){\line(1,0){4}}
        \put(85,16){\line(1,0){4}}
        \put(92,16){\line(1,0){4}}

        \put(85,0){\line(1,0){4}}
        \put(89,8){\line(1,0){3}}
        \put(92,0){\line(1,0){4}}

        \put(85,0){\line(0,1){16}}
        \put(96,0){\line(0,1){16}}
        \put(79,0){\line(0,1){16}}
        \put(80,0){\line(0,1){16}}
        \put(89,0){\line(0,1){16}}
        \put(92,0){\line(0,1){16}}
        \put(79,16){\oval(8,32)[bl]}
        \put(80,16){\oval(8,32)[br]}

        \end{picture}
        \par\vskip-6.5mm
        \thicklines}

\border\headpic {\hbox to\hsize{
\vbox{\noindent ITP--UH - 01/95 \hfill January 1995  \\
UMDEPP 95--90  \hfill hep-th/9501140 \\ }}}

\noindent
\vskip1.0cm

\begin{center}
{\Large\bf NO $~N=4~$ STRINGS ON WOLF SPACES~\footnote{
Supported in part by the `Deutsche Forschungsgemeinschaft',
\newline ${~~~~~}$ the NATO Grant CRG 930789, and the US National
Science Foundation Grant PHY-91-19746}}\\
\vglue.3in

S. James Gates, Jr.

{\it Department of Physics, University of Maryland, College Park,
MD 20742, USA}\\
{\sl gates@umdhep.umd.edu} \\
%\vglue.1in
and\\
%\vglue.1in
Sergei V. Ketov \footnote{
On leave of absence from:
High Current Electronics Institute of the Russian Academy of Sciences,
\newline ${~~~~~}$ Siberian Branch, Akademichesky~4, Tomsk 634055, Russia}

{\it Institut f\"ur Theoretische Physik, Universit\"at Hannover}\\
{\it Appelstra\ss{}e 2, 30167 Hannover, Germany}\\
{\sl ketov@itp.uni-hannover.de}
\end{center}
\vglue.1in
\begin{center}
{\Large\bf Abstract}
\end{center}
We generalize the standard $N=2$ supersymmetric Kazama-Suzuki coset
construction to the $N=4$ case by requiring the {\it non-linear}
(Goddard-Schwimmer) $N=4~$ quasi-superconformal algebra to be realized on
cosets. The constraints that we find allow very simple geometrical
interpretation and have the Wolf spaces as their natural solutions. Our results
 obtained by using components-level superconformal field theory methods are
fully consistent with standard results about $N=4$ supersymmetric
two-dimensional non-linear sigma-models and $N=4$ WZNW models on Wolf
spaces. We construct the actions for the latter and express the quaternionic
structure, appearing in the $N=4$ coset solution, in terms of the symplectic
structure associated with the underlying Freudenthal triple system. Next, we
gauge the $N=4~$ QSCA and build a quantum BRST charge for the $N=4$ string
propagating on a Wolf space. Surprisingly, the BRST charge nilpotency
conditions rule out the non-trivial Wolf spaces as consistent string
backgrounds.

% ======================= END of TITLE PAGE =============================

\newpage

% =======================================================================

\section{Introduction}

The critical (non-topological)~\footnote{By non-topological strings we mean
strings based on untwisted {\it two-dimensional} (2d) \newline ${~~~~~}$
(super)conformal algebras, with the usual relation between spin and
statistics.} $N=4$ strings are known since 1976 \cite{ade}, but they received
little attention in the literature because
of their apparently `negative' critical dimension. By the critical dimension
one actually means the formal number of irreducible 2d scalar $N=4$ multiplets
whose contribution to the conformal anomaly cancels the contribution of $N=4$
ghosts that arise in gauge-fixing the $N=4$ superconformal supergravity
multiplet. A closer inspection of the argument reveals at least two relevant
things: (i) it is implicit that the $N=4$ string constraints have to form  the
 `{\it small}\/' linear $N=4$ {\it superconformal algebra} (SCA) having the
$\Hat{su(2)}$ affine Lie subalgebra, and (ii) the background space in which
such $N=4$ strings are supposed to propagate is {\it flat}.

In this paper, we are going to challenge both assumptions in an attempt to find
 new consistent $N=4$ string theories. First af all, we replace the
 `small' linear $N=4$ SCA by the more general non-linear $N=4$ {\it
quasi-superconformal algebra} (QSCA) found by Goddard and Schwimmer \cite{gs}
and closely related with the `large' linear $N=4$ SCA, having two affine
$\Hat{su(2)}$ subalgebras. Second, we choose a {\it coset\/} $G/H$ as the
embedding space. The embedding space should be general enough to accomodate as
much as possible representations of the underlying QSCA, but not to be too
general in order to still allow an explicit treatment. Cosets perfectly
satisfy both requirements, as is well known in {\it (super)conformal field
theory} (SCFT). Requiring $N=4$ supersymmetry severely constrains the cosets
in question, and it is one of our main purposes to determine which cosets are
compatible with the $N=4~$ non-linear QSCA.

We first generalize the standard $N=0,1$ {\it Goddard-Kent-Olive} (GKO)
\cite{gko} and $N=2$ {\it Kazama-Suzuki} (KS) \cite{ks} coset constructions
 to the $N=4$ case (sects.~2 and 3). Next, we require the $N=4$ supersymmetry
in the general 2d non-linear scalar field theory and in the
{\it Wess-Zumino-Witten-Novikov} (WZNW) models (sect.~4), which complements
the $N=4~$ SCFT construction of sect.~3. As far as the {\it linear} $N=4~$
SCA's are concerned, Sevrin and Theodoridis \cite{st} found an
$N=4$ generalization of the GKO and KS coset constructions in SCFT by imposing
 the `large' linear $N=4~$ SCA in $N=1$ superspace. They found coset solutions
of the type $W\otimes SU(2)\otimes U(1)$, where $W$ is a Wolf space. We take a
 different approach by requiring a coset to support the non-linear $N=4$ QSCA,
 and using components. Our constraints allow very simple geometrical
interpretation, and have just the Wolf spaces as their solutions. Our SCFT
results are perfectly consistent with the standard results about the 2d {\it
non-linear sigma-models} (NLSM's) with $N$-extended supersymmetry. To solve
our $N=4$ constraints completely, we provide their alternative derivation, by
 constructing the relevant $N=4$ WZNW models on Wolf spaces. Based on the
triple system construction of the $N$-extended SCA's developed by G\"unaydin
\cite{gu}, we express the quaternionic structure, appearing in the $N=4$ coset
 solution, in terms of the symplectic structure associated with the underlying
 {\it Freudenthal triple system} (FTS). Next, we promote the symmetry realized
 by the $N=4~$ QSCA to the local level in order to get the corresponding $N=4$
string, and build the string BRST charge. Requiring its nilpotency is shown to
lead to severe constraints on the cosets in question. Finally, we briefly
discuss a connection to the known results \cite{bw,pn} about the on- and
off-shell structure of matter couplings in extended supergravities in four and
two dimensions (sect.~5). Our conclusion and outlook are summarized in sect.~6.
 The defining equatons of the $N=4~$ QSCA are collected in Appendix.
\vglue.2in

\section{Supersymmetric Coset Constructions}

In this section we review some well-known standard constructions in 2d SCFT,
including the KS construction for $N=2$. This gives the necessary pre-requisite
 for the $N=4~$ SCFT coset construction to be discussed in the next section,
and introduces our notation.
\vglue.2in

\subsection{Affine Lie algebras and Sugawara construction}

Let $\cg$ be the Lie algebra associated with a semi-simple Lie group $G$, and
$f^{abc}$ and $|G|$ be its structure constants and dimension,
respectively, $a,b=1,2,\ldots,|G| $. Given a non-trivial representation
$t^a_{(r)}$ of $\cg$, let us consider the trace,
$\tr\left(t^a_{(r)}t^b_{(r)}\right)\equiv g^{ab}_{(r)}$, defining the
normalization metric $g^{ab}_{(r)}$. This metric can always be diagonalized in
 the representation space ($\cg$-module),
$$\tr\left(t^a_{(r)}t^b_{(r)}\right)= l_r\d^{ab}~.\eqno(2.1)$$
In particular, as far as the adjoint $(A)$ representation is concerned, the
metric $g^{ab}_A$ is known as the Cartan-Killing metric, and its canonical
form is given by
$$f^{acd}f^{bdc} = l_A\d^{ab}~.\eqno(2.2)$$

The Casimir eigenvalue $C_r$ associated with representation $t^a_{(r)}$ is
defined by
$$C_r\d^{\a\b}=\sum_a \left(t^a_{(r)}t^a_{(r)}\right)^{\a\b}~.\eqno(2.3)$$
Eqs.~(2.1) and (2.3) imply the relation $ C_r d_r = l_r |G|~,$
where the dimension $d_r$ of representation $(r)$ has been
introduced, $\a,\b=1,2,\ldots, d_r$. The normalization of representation $(r)$
is therefore fixed by the coefficient $l_r$ alone. If the sum in eq.~(2.1)
were restricted to the Cartan subalgebra of $\cg$, we would get instead
$$ \sum^{d_r}_{k=1} \m^2_{(k)} = l_r r_G~,\eqno(2.4a)$$
where $r_G$ is the rank of the group $G$, and $\m$ are the weights of the
representation $(r)$. In particular, as far as the adjoint representation
is concerned, we have $d_A=|G|$ and
$$C_A=l_A= r_G^{-1}\sum^{|G|}_{a=1}\a^2_{(a)}~,\eqno(2.4b)$$
where $\a$'s are the roots of $\cg$. Let $\j$ be the highest root. Then the
normalization-independent quantity
$$\tilde{h}_G\equiv C_A/\j^2=\fracmm{1}{r_G}\left[ n_L +
\left(\fracmm{S}{L}\right)^2 n_S\right]~,\eqno(2.5)$$
where $n_L$ and $n_S$ denote the numbers of long and short roots, respectively,
 is known as the dual Coxeter number. The roots in classical Lie algebras are
known to come in two lengths at most. The Dynkin diagrams having only
single lines have roots all of the same length, and they correspond to the
so-called simply-laced Lie algebras.

Let $J^a(z)$ be generators for the associated affine Lie algebra $\Hat{\cg}$
of level $k_G$,
$$J^a(z)J^b(w)\sim {\d^{ab}k_G/2\over (z-w)^2} + {if^{abc}\over z-w}
J^c(w)~. \eqno(2.6) $$
The Sugawara stress tensor is defined by~\footnote{Normal ordering is implicit
 in our formulae.}
$$T(z) = \fracmm{1}{k_G + \tilde{h}_G}\sum_{a=1}^{|G|} J^a(z)J^a(z) ~,
\eqno(2.7) $$
and it has central charge
$$ c_G = \fracmm{k_G|G|}{k_G + \tilde{h}_G}~. \eqno(2.8) $$

One can think of this CFT construction as realized by the 2d WZNW theory based
on the group $G$ (see sect.~4 for more). As is well known, the level $k_G$ must
 be a positive integer for unitary affine representations, as well as for the
WZNW action to be well-defined.
\vglue.2in

\subsection{Super-affine Lie algebras and the associated
super-Virasoro algebras}

The WZNW theory is the particular 2d non-linear sigma-model (with WZ torsion)
on a group manifold, and it can be made $(N=1)$ supersymmetric along the
standard lines, either in components or in superspace. It follows that the
WZNW fermions, which are the superpartners of the WZNW bosons (in the adjoint
representation), are actually {\it free} fields (sect.~4). This can be
understood by noticing that the WZNW fields take their values in a group
manifold with the {\it parallelizing} torsion represented by the WZ term and,
hence, the spin connection present in the Lie algebra-valued covariant
derivative acting on the WZNW fermions should be trivial.

Let $\j^a(z)$ be a set of (holomorphic) free fermions in the adjoint
representation, which can be thought of as originated from the super-WZNW
theory, with the canonical OPE's
$$\j^a(z)\j^b(w)\sim - {\d^{ab}\over z-w}~.\eqno(2.9) $$
One can always associate affine currents with free fermions,
$$J^a_f(z)={i\over 2}f^{abc}\j^b(z)\j^c(z)~,\eqno(2.10) $$
which define a representation of $\Hat{\cg}$ at level $k_G=\tilde{h}_G$. The
Sugawara construction for free fermions in the adjoint representation gives
the stress tensor which is equivalent to the usual free (quadratric in the
fields) fermionic stress tensor, and it has the central charge $c_f=\ha |G|$,
as it should.

This is to be compared with the defining OPE's of an $N=1$ supersymmetric
 affine Lie algebra,
$$\eqalign{
J^a(z)J^b(w) ~\sim~ & {\d^{ab}k_G/2\over (z-w)^2}+{if^{abc}J^c(w)\over z-w}~,
\cr
J^a(z)j^b(w) ~\sim~ & {if^{abc}j^c(w)\over z-w}~,\cr
j^a(z)j^b(w) ~\sim~ & {\d^{ab}k_G/2\over z-w}~.}\eqno(2.11) $$
Defining
$$j^a(z)=j^a_f(z)\equiv i\sqrt{\tilde{h}_G/2}\,\j^a(z)~,\eqno(2.12a) $$
and
$$J^a(z)=J^a_f(z)\equiv -{i\over\tilde{h}_G}f^{abc}j^b(z)j^c(z)~,
\eqno(2.12b) $$
we therefore obtain the free-fermionic representation of the super-affine Lie
algebra at the level $k_G=\tilde{h}_G$. Similarly, the Sugawara bosonic
construction can also be supersymmetrized to the full  $N=1$ super-Virasoro
algebra by introducing a dimension-3/2 current $G_f(z)$ which is the
superpartner of $T_f(z)$. The supercurrent $G_f$ must square to $T_f$,
and its explicit form is given by
$$G_f(z)=-{1\over 3\sqrt{2\tilde{h}_G}}f^{abc}\j^a\j^b\j^c~.\eqno(2.13) $$
One easily finds
$$\eqalign{
G_f(z)J^a_f(w) ~\sim~ & {1\over (z-w)^2}j^a_f(z)~,\cr
G_f(z)j^a_f(w) ~\sim~ & {1\over z-w}J^a_f(w)~.}\eqno(2.14) $$

Of course, all the above-mentioned is valid for {\it any} free fermions, not
just for those belonging to the super-WZNW theory. If, nevertheless, our free
fermions originate from the super-WZNW theory, we still have at our disposal
the bosonic currents $\hat{J}^a(z)$ forming a level-$k_G$ representation of
affine Lie algebra $\Hat{\cg}$, which are independent on the fermionic fields.
We are therefore in a position to define general affine representations,
$$J^a(z)= J^a_f(z) + \hat{J}^a(z)~, \eqno(2.15) $$
of level
$$k=k_G +\tilde{h}_G~,\eqno(2.16) $$
and of central charge
$$c={k_G|G| \over k_G + \tilde{h}_G} + {\frac 12}|G|~.\eqno(2.17) $$

It can be extended to a representation of the super-affine algebra by adding
$$j^a(z)=i\sqrt{k/2}\,\j^a(z)~.\eqno(2.18) $$

The (Sugawara-type) $N=1$ super-Virasoro algebra associated with this
construction is given by
$$\eqalign{
T(z)=~&~ {1\over k}\left[ \hat{J}^a(z)\hat{J}^a(z) - j^a(z)\pa j^a(z)
\right]~,\cr
G(z)=~&~ {2\over k}\left[ j^a(z)\hat{J}^a(z) - {i\over 3k}f^{abc}j^a(z)
j^b(z)j^c(z)\right]~.\cr} \eqno(2.19) $$

The 2d field theory realization of this CFT construction is provided by the
quantized super-WZNW theories (sect.~4).
\vglue.2in

\subsection{Coset (GKO) constructions}

A much larger class of (S)CFT's can be obtained by the coset method, also known
as the GKO construction. It was even conjectured that coset models may exhaust
all {\it rational} conformal field theories (RCFT's). Let $H$ be a subgroup of
 $G$, $\ch$ the Lie algebra of $H$, and $J^i_H(z)$ the affine
$\hat{\ch}$-currents, $i,j=1,2,\ldots,|H|$. We assume that the first $|H|$
currents in $\{J^a_G\}$ just represent the currents $\{J^i_H\}$. As far as our
 notation is concerned, {\it early} lower  case Latin
indices are used for $G$-indices, {\it middle} lower case Latin indices are
used for $H$-indices, while early lower case Latin indices {\it with bars} are
used for $G/H$-indices, $a=(i,\bar{a})$ and $\bar{a}=|H|+1,\ldots,|G|$. We have
$$\eqalign{
J^a_G(z)J^b_G(w)\sim & {\d^{ab}k_G/2\over (z-w)^2} + {if^{abc}\over z-w}
J^c_G(w)~,\cr
J^i_H(z)J^j_H(w)\sim & {\d^{ij}k_H/2\over (z-w)^2} + {if^{ijk}\over z-w}
J^k_H(w)~.}\eqno(2.20)$$

The level $k_H$ is determined by embedding of $\ch$ into $\cg$. An embedding
is characterized by the embedding index $I_H$ defined by $I_H=\j^2_G/\j^2_H$
which is always an integer. As far as the bosonic WZNW currents
$\{\hat{J}_H\}\subset\{\hat{J}_G\}$ are concerned, we obviously have
$k_H=I_H k_G$. In particular, if the simple roots of $H$ form a subset of the
simple roots of $G$, then $I_H=1$ and $k_H=k_G$.~\footnote{$I_H=3$ when
$G=G_2$ and $H=SU(2)$, whereas $I_H=2$ when $G=SO(7)$ and $H=SO(3)$.}

Having restricted the free fermions $\j^a$ in the adjoint of $G$ to the subset
$\j^i$ in the adjoint of $H$, we can introduce the affine currents
$$J^i_{H,f}(z)=\fracmm{i}{2}f^{ijk}\j^j(z)\j^k(z) \eqno(2.21)$$
forming a representation of $\hat{\ch}$ of level $k_H=\tilde{h}_H$. Still,
there is another natural representation of $\hat{\ch}$, also associated with
the free fermions and defined by the currents
$$J^i_{G/H,f}(z)=\fracmm{i}{2}f\ud{i}{\bar{b}\bar{c}}\j^{\bar{b}}(z)
\j^{\bar{c}}(z) \eqno(2.22)$$
of level $k_{H,f}=I_H\tilde{h}_G -\tilde{h}_H$, where $\tilde{h}_H$ is the dual
Coxeter number for $\ch$. Therefore, after taking into account eqs.~(2.15) and
(2.16), one finds that $k_H=I_Hk_G + I_H\tilde{h}_G-\tilde{h}_H$  in general.

The Sugawara stress tensors associated with the affine $G$ and $H$ currents
take the form
$$\eqalign{
T_G(z)= & {1\over k_G + \tilde{h}_G}\sum_{a=1}^{|G|} J^a_G(z)J^a_G(z)~,\cr
T_H(z)= & {1\over k_H + \tilde{h}_H}\sum_{i=1}^{|H|} J^i_H(z)J^i_H(z)~.}
\eqno(2.23) $$
The corresponding Virasoro central charges are
$$c_G = {k_G|G|\over k_G + \tilde{h}_G}~,\quad
  c_H = {k_H|H|\over k_H + \tilde{h}_H}~.\eqno(2.24) $$

The standard (GKO) coset construction is defined by
$$T_{G/H}= T_G - T_H~,\eqno(2.25a) $$
and it has  central charge
$$c_{G/H}=c_G - c_H = {k_G|G|\over k_G + \tilde{h}_G} - {k_H |H|\over k_H +
\tilde{h}_H}~.\eqno(2.25b) $$

The $N=1$ generalization of the GKO construction given above is based on an
{\it orthogonal} decomposion of the $N=1$ SCA associated with the group $G$,
with respect to its subgroup $H$,
$$\eqalign{
T_G(z) = & T_H(z) + T_{G/H}(z)~,\cr
G_G(z) = & G_H(z) + G_{G/H}(z)~,\cr}\eqno(2.26) $$
where $H$- and $G/H$- currents are to be mutually commuting. To actually get
such a decomposition, one uses two $\hat{\ch}$-representations introduced in
  eqs.~(2.12), (2.15) and (2.22), namely
$$\tilde{J}^i(z) = \hat{J}^i(z) + {i\over 2}f^{i\bar{b}\bar{c}}\j^{\bar{b}}
(z)\j^{\bar{c}}(z)~.\eqno(2.27) $$
and
$$J^i(z)=\tilde{J}^i(z) + {i\over 2}f^{imn}\j^m(z)\j^n(z)~,\eqno(2.28) $$
where $\{\hat{J}_H\}\subset\{\hat{J}_G\}$ are the bosonic currents forming a
level-$k_H$ representation of $\hat{\ch}$. The stress tensor $T_H(z)$
and the supercurrent $G_H(z)$ are defined by
$$\eqalign{
T_H(z)=~& {1\over k}\tilde{J}^i(z)\tilde{J}^i(z) + \ha\j^i(z)\pa\j^i(z)~,\cr
G_H(z)=~& {i\over\sqrt{2k}}\j^i(z)\tilde{J}^i(z) - {1\over 3\sqrt{2k}}
f^{imn}\j^i(z)\j^m(z)\j^n(z)~,\cr}\eqno(2.29)  $$
where eq.~(2.19) has been used as a guide. Note that the `improved' current
$\tilde{J}^i$ instead of the `naive' bosonic current $\hat{J}^i$  has been
used in eq.~(2.29). This is possible since $\tilde{J}^i(z)$ commutes with
$j^i(z)$. Most importantly, eq.~(2.26) yields the desired {\it orthogonal}
decomposition since so {\it defined} $T_{G/H}(z)$ and  $G_{G/H}(z)$ commute
with $J^i(z)$, $\tilde{J}^i(z)$ and $j^i(z)$.~\footnote{The generators $T_G$
and $G_G$ have been introduced before, see eq.~(2.19).}  Explicitly, they read
$$\eqalign{
T_{G/H}(z) =~ &  \fracmm{1}{k}\left[ \hat{J}^{\bar{a}}(z)\hat{J}^{\bar{a}}(z)
+ \fracmm{k_G}{2}\j^{\bar{a}}(z)\pa\j^{\bar{a}}(z)
- i\hat{J}^i(z)f^{i\bar{b}\bar{c}}\j^{\bar{b}}(z)\j^{\bar{c}}(z)
\right. \cr
& \left. + f^{\bar{a}\bar{c}\bar{d}}f^{\bar{b}\bar{c}\bar{d}}
\j^{\bar{a}}(z)\pa\j^{\bar{b}}(z) -
\fracmm{1}{4}f^{\bar{a}\bar{b}\bar{c}}f^{\bar{a}\bar{g}\bar{f}}\j^{\bar{b}}
(z)\j^{\bar{c}}(z)\j^{\bar{g}}(z)\j^{\bar{f}}(z)\right]~,\cr
G_{G/H}(z) =~ & \fracmm{i}{\sqrt{2k}} \j^{\bar{a}}(z)\hat{J}^{\bar{a}}(z)
-\fracmm{1}{3\sqrt{2k}}f^{\bar{a}\bar{b}\bar{c}}\j^{\bar{a}}(z)\j^{\bar{b}}(z)
\j^{\bar{c}}(z)~,\cr}\eqno(2.30)$$
and have central charge
$$ c_{G/H}=c_G - c_H = \left[ \fracmm{k_G|G|}{k_G+\tilde{h}_G}+\ha|G|\right]
 - \left[ \fracmm{(I_Hk_G+I_H\tilde{h}_G-\tilde{h}_H)|H|}{I_H(k_G+\tilde{h}_G)}
+\ha|H|\right]~;$$
$$ c_{G/H}= {3k_G\over 2k}\dim(G/H)+\fracmm{1}{8k}f^{\bar{a}\bar{b}\bar{c}}
f^{\bar{a}\bar{b}\bar{c}}~,\qquad {\rm when}\quad I_H=1~,\eqno(2.31) $$
where $k=k_G+\tilde{h}_G\,$ as above. In particular, for a {\it symmetric}
space $G/H$ where $f^{\bar{a}\bar{b}\bar{c}}=0$, one finds
$$\eqalign{
T_{G/H}(z)~=~ & {1\over k}\left[ \hat{J}^{\bar{a}}(z)\hat{J}^{\bar{a}}(z) +
{k_G\over 2}\j^{\bar{a}}(z)\pa\j^{\bar{a}}(z) - if^{i\bar{b}\bar{c}}
\hat{J}^i(z)\j^{\bar{b}}(z)\j^{\bar{c}}(z)\right]~,\cr
G_{G/H}(z)~=~ & {i\over\sqrt{2k}}\j^{\bar{a}}(z)\hat{J}^{\bar{a}}(z)~,}
\eqno(2.32)$$
and
$$c_{G/H}=c_G - c_H ={3k_G\over 2(k_G+\tilde{h}_G)}\dim(G/H)~,\eqno(2.33) $$
according to eqs.~(2.16) and (2.31) with $I_H=1$.
\vglue.2in

\subsection{KS construction}

Having obtained the $N=1$ super-Virasoro algebra associated with the $N=1$
super affine Lie algebra, it is quite natural to ask about the conditions on
the coset $G/H$ which would allow more supersymmetries, i.e. $N>1$. The case
of $N=2$ was fully addressed by Kazama and Suzuki \cite{ks}. Since the
$N=2$ extended SCA has a second supercurrent and an abelian $U(1)$ current
beyond the content of the $N=1~$ SCA, the $N=2$ conditions on the coset $G/H$
just originate from requiring their existence. The most general ansatz for the
second supercurrent takes the form \cite{ks}
$$G^{(2)}(z)=\fracmm{i}{\sqrt{2k}}h_{\bar{a}\bar{b}}\j^{\bar{a}}(z)
\hat{J}^{\bar{b}}(z) -\fracmm{1}{3\sqrt{2k}}S^{\bar{a}\bar{b}\bar{c}}
\j^{\bar{a}}(z)\j^{\bar{b}}(z)\j^{\bar{c}}(z)~,\eqno(2.34)$$
where $h_{\bar{a}\bar{b}}$ and $S^{\bar{a}\bar{b}\bar{c}}$ are constants.
The supercurrents $G^{(1)}\equiv G_{G/H}$ and $G^{(2)}$ have to satisfy the
basic $N=2~$ SCA~ OPE
$$G^{(i)}(z)G^{(j)}(w)~\sim~ \fracmm{2\d^{ij}c/3}{(z-w)^3} +
\fracmm{2i\ve^{ij}J(w)}{(z-w)^2}+\fracmm{2\d^{ij}T(w)+i\d^{ij}\pa J(w)}{z-w}~,
\eqno(2.35)$$
where the $N=2~$ SCA current $J(z)$ has been introduced. It results in the
following $N=2$ conditions \cite{ks}:
$$ \eqalign{
(i) ~ & ~~~ h_{\bar{a}\bar{b}} = -h_{\bar{b}\bar{a}}~,\quad
h_{\bar{a}\bar{c}}h_{\bar{c}\bar{b}}=-\d_{\bar{a}\bar{b}}~,\cr
(ii) ~ & ~~~ h_{\bar{a}\bar{c}}f_{\bar{c}\bar{b}d}=
h_{\bar{b}\bar{c}}f_{\bar{c}\bar{a}d}~,\cr
(iii) ~ & ~~~f_{\bar{a}\bar{b}\bar{c}}=h_{\bar{a}\bar{f}}h_{\bar{b}\bar{g}}
f_{\bar{f}\bar{g}\bar{c}} + h_{\bar{b}\bar{f}}h_{\bar{c}\bar{g}}
f_{\bar{f}\bar{g}\bar{a}} + h_{\bar{c}\bar{f}}h_{\bar{a}\bar{g}}
f_{\bar{f}\bar{g}\bar{b}}~,\cr
(iv) ~ & ~~~S_{\bar{a}\bar{b}\bar{c}} = h_{\bar{a}\bar{f}}h_{\bar{b}\bar{g}}
h_{\bar{c}\bar{h}}f_{\bar{f}\bar{g}\bar{h}}~,\cr}\eqno(2.36)$$
Given these conditions, the $N=2~$ SCA~ $U(1)$ current reads
$$J(z) = -\fracmm{i}{\sqrt{2}}h_{\bar{a}\bar{b}}\j^{\bar{a}}(z)
\j^{\bar{b}}(z) +\fracmm{1}{k}h_{\bar{c}\bar{d}}f^{\bar{c}\bar{d}a}
\left[\hat{J}^a(z)+\fracmm{i}{2}f^{a\bar{a}\bar{b}}\j^{\bar{a}}(z)
\j^{\bar{b}}(z)\right]~.\eqno(2.37)$$

The conditions (2.36) have simple geometrical interpretation, which allows to
describe their solutions in full \cite{ks}. In particular, the condition (i)
just means that $h_{\bar{a}\bar{b}}$ is an {\it almost complex} structure on a
{\it hermitian} manifold. The condition (ii) implies that the almost complex
structure is {\it covariantly} constant with respect to the connection with
torsion to be defined by the structure constants, whereas the condition (iii)
means that the almost complex structure is {\it integrable}, i.e. it is a
{\it complex structure} indeed (the equation (iii) is equivalent to the
vanishing condition on the so-called Nijenhuis tensor \cite{ghr}). The
condition (iv) is the defining equation for $S_{\bar{a}\bar{b}\bar{c}}$.  The
conditions (ii), (iii) and (iv) are trivially satisfied for the {\it symmetric}
 spaces having $f^{\bar{a}\bar{b}\bar{c}}=S^{\bar{a}\bar{b}\bar{c}}=0$. The
{\it hermitian symmetric} spaces therefore represent an important class of
solutions to eq.~(2.36), and they were extensively studied \cite{ks}. A
different class of $N=2$ supersymmetric solutions is given by the
{\it k\"ahlerian} coset spaces which are in fact the {\it only\/} solutions if
${\rm rank}\,G={\rm rank}\,H$ \cite{ks}. In general, when
${\rm rank}\,G \, - \, {\rm rank}\,H=2n$, $n=0,1,2,\ldots \/$, the coset
$G/ \left[H\otimes U(1)^{2n}\right]$ must be k\"ahlerian \cite{ks}. Hence,
a solution to the $N=2$ conditions exists for {\it any hermitian} coset space.
Given a Cartan-Weyl decomposition of $\cg$, the complex structure maps
the Cartan subalgebra of $\cg$ into itself, whereas the generators
corresponding to positive (negative) roots are the eigenvectors with the
eigenvalues $+i$ $(-i)$.
\vglue.2in

\section{$N=4~$ SCFT coset models}

The KS construction delivers a large class of $N=2~$ SCFT's by the coset space
method. We now wish to identify those of them which actually possess $N=4$
supersymmetry. Sevrin and Theodoridis \cite{st} already generalized the KS
construction to the $N=4$ case by requiring the existence of the `large'
{\it linear} $N=4$ SCA having $D(2,1;\a)$ as projective subalgebra. The $N=4$
generators are supposed to act on a coset $G/H$, i.e. they have to commute
with the $H$ generators. Our approach to constructing $N=4~$ SCFT's by the
coset space method is however different from the one adopted in ref.~\cite{st}.
 We are going to impose the {\it non-linear} $N=4$ supersymmetry because it is
more general than the linear one represented by the `large' $N=4~$ SCA.
The `large' linear $N=4~$ SCA is actually {\it not} a symmetry algebra since it
 has subcanonical charges represented by four free fermions and one boson.
The proper $N=4$ supersymmetric symmetry algebra having only canonical charges
 of dimension 2, 3/2 and 1 was constructed by Goddard and Schwimmer \cite{gs},
and we are going to call it the $\hat{D}(2,1;\a)$ {\it quasi-superconformal
algebra} (QSCA) \cite{k4}. The $N=4~$ QSCA $\hat{D}(2,1;\a)$ is quadratically
{\it non-linearly} generated.  Given a SCFT representing the `large' linear
$N=4~$ SCA, one can always realize over there the $\hat{D}(2,1;\a)$ QSCA too,
since the generators of the latter can be non-linearly constructed from the
generators of the former (see Appendix). The reverse may not be always
possible. We should therefore expect more solutions to exist when imposing the
$\hat{D}(2,1;\a)$ QSCA instead of the `large' linear $N=4~$ SCA. In addition,
imposing the QSCA seems to be more satisfactory from the viewpoint of $N=4$
string theory: the most general algebra to be gauged is not the `large' linear
 $N=4~$ SCA but the non-linear $\hat{D}(2,1;\a)$ QSCA! \footnote{See Appendix
for a review of both algebras.}

The $\hat{D}(2,1;\a)$ QSCA comprises stress tensor $T(z)$, four dimension-3/2
supercurrents $G^{\m}(z)$, and six dimension-1 currents $J^{\m\n}(z)$ in the
adjoint of $SO(4)\cong SU(2)_+\otimes SU(2)_-$. The only non-trivial OPE of
this QSCA defines an $N=4$ supersymmetry algebra in the form
$$\eqalign{
G^{\m}(z)G^{\n}(w)~\sim~~~ &~\fracmm{4k^+k^-}{(k^+ + k^-+2)}
\fracmm{\d^{\m\n}}{(z-w)^3}+\fracmm{2T(w)\d^{\m\n}}{z-w}\cr
&  - \fracmm{k^+ + k^-}{k^+ + k^- +2}
\left[\fracmm{2J^{\m\n}(w)}{(z-w)^2} +\fracmm{\pa J^{\m\n}(w)}{z-w}\right]\cr
& + \fracmm{k^+ - k^-}{k^+ + k^- +2}\ve^{\m\n\r\l}
\left[\fracmm{J^{\r\l}(w)}{(z-w)^2} +
\fracmm{\pa J^{\r\l}(w)}{2(z-w)}\right] \cr
& - \fracmm{\ve^{\m\r\l\z}\ve^{\n\r\h\o}}{2(k^+ + k^- +2)}
\fracmm{:J^{\l\z}J^{\h\o}:(w)}{(z-w)}~,\cr} \eqno(3.1)$$
where $k^+$ and  $k^-$ are levels of affine Lie algebras associated with
$SU(2)_+$ and $SU(2)_-$, respectively. The tensor $J^{\m\n}$ comprises two
(anti)self-dual $SU(2)$ triplets $(M=1,2,3)$
$$J^{\m\n}(z)=(t^{M-})^{\m\n}J^{M-}(z) +(t^{M+})^{\m\n}J^{M+}(z)~,\eqno(3.2)$$
where the antisymmetric $4\times 4$ matrices $t^{M\pm}$ satisfy the relations
$$\[t^{M\pm},t^{N\pm}\]=-2\ve^{MNP}t^{P\pm}~,\quad \[t^{M+},t^{N-}\]=0~,\quad
\{t^{M\pm},t^{N\pm}\}=-2\d^{MN}~.\eqno(3.3)$$

The only non-linear $:JJ:(w)$ term on the r.h.s. of eq.~(3.1) can be
interpreted as the Sugawara stress tensor for the $\Hat{SO(4)}$ currents. It
 attributes the $N=4$ `improvement' to the `naive' stress tensor $T(z)$.

Requiring the $N=4~$ QSCA supersymmetry, we expect the KS conditions (2.36)
to be satisfied for each supersymmetry separately. This happens to be true
indeed (see below). On dimensional grounds, the general ansatz (2.34) is valid
 for {\it any} supersymmetry,
$$G^{\m}(z)=\fracmm{i}{\sqrt{2k}}\left[ h^{\m}_{\bar{a}\bar{b}}\j^{\bar{a}}(z)
\hat{J}^{\bar{b}}(z) + \fracmm{i}{3}S^{\m}_{\bar{a}\bar{b}\bar{c}}
\j^{\bar{a}}(z)\j^{\bar{b}}(z)\j^{\bar{c}}(z)\right]~,\eqno(3.4)$$
where $h^{\m}_{\bar{a}\bar{b}}$ and $S^{\m}_{\bar{a}\bar{b}\bar{c}}$ are
constants, $\m=0,1,2,3$. The OPE for a product of the supercurrents (3.4)
takes the form
$$\eqalign{
G^{\m}(z)G^{\n}(w)~\sim~&~ -\fracmm{1}{2k}\left\{
\fracmm{1}{(z-w)^3}\left[
-\fracmm{k_G}{2}h^{\m}_{\bar{a}\bar{b}}h^{\n}_{\bar{a}\bar{b}}
-\fracmm{1}{9}S^{\m}_{\bar{a}\bar{b}\bar{c}}S^{\n}_{\bar{a}\bar{b}\bar{c}}
\right]\right.\cr
{}~&~+\fracmm{1}{(z-w)^2}\left[
\fracmm{k_G}{2}h^{\m}_{\bar{a}\bar{c}}h^{\n}_{\bar{b}\bar{c}}\j^{\bar{a}}
\j^{\bar{b}} -ih^{\m}_{\bar{a}\bar{c}}h^{\n}_{\bar{a}\bar{g}}
f^{\bar{c}\bar{g}d}\hat{J}^d
+\fracmm{1}{3}S^{\m}_{\bar{a}\bar{b}\bar{f}}S^{\n}_{\bar{a}\bar{b}\bar{g}}
\j^{\bar{f}}\j^{\bar{g}} \right]\cr
{}~&~+\fracmm{1}{z-w}\left[ -h^{\m}_{\bar{c}\bar{a}}h^{\n}_{\bar{c}\bar{b}}
\hat{J}^{\bar{a}}\hat{J}^{\bar{b}} +ih^{\m}_{\bar{a}\bar{c}}h^{\n}_{\bar{b}
\bar{g}}f^{\bar{c}\bar{g}d}\hat{J}^d\j^{\bar{a}}\j^{\bar{b}}
-2ih^{(\m}_{\bar{a}\bar{b}}S^{\n)}_{\bar{a}\bar{g}\bar{h}}\hat{J}^{\bar{b}}
\j^{\bar{g}}\j^{\bar{h}} \right. \cr
{}~&~\left.\left.
-\fracmm{k_G}{2}h^{\m}_{\bar{b}\bar{c}}h^{\n}_{\bar{a}\bar{c}}
\j^{\bar{a}}\pa \j^{\bar{b}}
-\fracmm{1}{3}S^{\m}_{\bar{a}\bar{b}\bar{g}}S^{\n}_{\bar{a}\bar{b}\bar{f}}
\j^{\bar{f}}\pa\j^{\bar{g}}
-\fracmm{1}{3}S^{\m}_{\bar{a}\bar{b}\bar{c}}S^{\n}_{\bar{a}\bar{f}\bar{g}}
\j^{\bar{b}}\j^{\bar{c}}\j^{\bar{f}}\j^{\bar{g}} \right]\right\}~.\cr }
\eqno(3.5)$$

Eq.~(3.5) is to be compared with eq.~(3.1). To get $T=T_{G/H}$ of eq.~(2.30)
on the r.h.s. of eq.~(3.5), let us first look at the coefficients of the terms
 $(z-w)^{-1}\hat{J}\hat{J}$. This gives the first necessary condition
$$h^{\m}_{\bar{a}\bar{b}}h^{\n}_{\bar{a}\bar{c}}+
h^{\n}_{\bar{a}\bar{b}}h^{\m}_{\bar{a}\bar{c}}=2\d^{\m\n}\d_{\bar{b}\bar{c}}
 ~.\eqno(3.6)$$
The supercharge $G^0=G_{G/H}$ of the $N=1$ subalgebra is defined according
to the last line of eq.~(2.30), which implies
$$h^{0}_{\bar{a}\bar{b}}=\d_{\bar{a}\bar{b}}~,\qquad
S^{0}_{\bar{a}\bar{b}\bar{c}}=f_{\bar{a}\bar{b}\bar{c}} ~.\eqno(3.7)$$
Substituting eq.~(3.7) into eq.~(3.6) at $\m=M$ and $\n=0$ yields
$$h^{M}_{\bar{a}\bar{b}}=-h^{M}_{\bar{b}\bar{a}} ~,\eqno(3.8)$$
whereas taking $\m=\n=M$ yields
$$h^{M}_{\bar{a}\bar{c}}h^{M}_{\bar{c}\bar{b}}=-\d_{\bar{a}\bar{b}} ~,\qquad
{\rm (no~~ sum~~ over~~} M) ~.\eqno(3.9)$$
Eqs.~(3.8) and (3.9) mean that each $h^{M}_{\bar{a}\bar{b}}$ represents
an {\it almost complex hermitian structure}. Altogether, according to
eq.~(3.6), they represent an {\it almost quaternionic tri-hermitian} structure.

The terms of the form $(z-w)^{-1}\hat{J}\j\j$ in eq.~(3.5) have to deliver
the remaining terms in the stress tensor $T_{G/H}$ of eq.~(2.30),
in particular. We find that this necessarily implies the two conditions:
$$h^{\m}_{\bar{g}[ \bar{a}}S^{\n\bar{g}}_{\bar{b}\bar{c}]} +
h^{\n}_{\bar{g}[ \bar{a}}S^{\m\bar{g}}_{\bar{b}\bar{c} ]} = 2\d^{\m\n}
f_{\bar{a}\bar{b}\bar{c}}~, \eqno(3.10)$$
and
$$h^{\m}_{\bar{a}\bar{b}}f_{\bar{b}\bar{c}d}=
h^{\m}_{\bar{b}\bar{c}}f_{\bar{a}\bar{b}d}~.\eqno(3.11)$$
Equation (3.10) determines the tensor
$S^{\m}_{\bar{a}\bar{b}\bar{c}}$ as follows:
$$S^{\m}_{\bar{a}\bar{b}\bar{c}}=h^{\m\bar{g}}_{[ \bar{a}}f_{\bar{b}\bar{c}]
 \bar{g}}~,\eqno(3.12)$$
or $S^{M}_{\bar{a}\bar{b}\bar{c}}=h^{M\bar{g}}_{[ \bar{a}}
f_{\bar{b}\bar{c}] \bar{g}}$. Together with eq.~(3.11), it gives us the
second consistency condition
$$h^{\m\bar{b}}_{ [ \bar{a}}h^{\n\bar{g}}_{\bar{c}}f_{d]\bar{b}\bar{g}}
+h^{\n\bar{b}}_{[ \bar{a}}h^{\m\bar{g}}_{\bar{c}}f_{d]\bar{b}\bar{g}}
=2\d^{\m\n}f_{\bar{a}\bar{c}d}~.\eqno(3.13)$$

So far, we only required the relevant stress tensor to appear on the r.h.s. of
the supersymmetry algebra in eq.~(3.5), which resulted in the necessary
conditions (3.6) and (3.13) for the cosets in question.  These equations are
also contained in the set of $N=4$ conditions found by Sevrin and Theodoridis
in their work \cite{st}. It is not surprising since they are not sensitive to
the differences between the `large' linear $N=4~$ SCA and the non-linear QSCA.
\footnote{As is shown in the next section, the same conditions follow by
requiring the $(1,0)$ supersymmetric \newline ${~~~~~}$ 2d non-linear
sigma-model to possess $(4,0)$ supersymmetry.} These conditions are therefore
very general, and they
also have very clear geometrical interpretation \cite{ghr}. Namely, according
to eq.~(3.6), there should be three independent almost complex hermitian
structures satisfying the quaternionic algebra, thus defining an almost
quaternionic tri-hermitian structure on $G/H$. Eqs.~(3.11) and (3.13)
guarantee the $H$-invariance and the covariant constancy of that structure,
and imply the vanishing of the Nijenhuis tensor \cite{ghr}. In other words,
the almost quaternionic structure is actually integrable, and defines a
quaternionic tri-hermitian structure. The latter appears to be the {\it only}
condition to be satisfied in order that a coset $G/H$ could support $N=4~$
SCFT. All quaternionic manifolds are known to be Einstein spaces of constant
non-vanishing scalar curvature. The only known {\it compact} cases are the
Wolf spaces to be discussed below.

Looking at the double-pole terms in eq.~(3.5) and comparing them with
eq.~(3.1), we find the $\Hat{SU(2)}_{\pm}$ currents of the QSCA in the form
$$J^{M-}(z)=\fracmm{1}{16}h^M_{\bar{a}\bar{b}}\j^{\bar{a}}(z)\j^{\bar{b}}(z)~,
\eqno(3.14)$$
and
$$J^{M+}(z)=\fracmm{i}{4(\tilde{h}_G-2)}\left[
h^M_{\bar{a}\bar{c}}f^{\bar{a}\bar{c}d}\hat{J}^d(z)
+\fracmm{1}{3}h^M_{\bar{g}[ \bar{a} }f_{\bar{b}\bar{c}]\bar{g}}
f_{\bar{a}\bar{b}\bar{f}}\j^{\bar{c}}(z)\j^{\bar{f}}(z) \right]~,\eqno(3.15)$$
which generalize the results of ref.~\cite{vp} to non-symmetric spaces.
Simultaneously, the levels of the affine Lie subalgebras
$\Hat{SU(2)}_{k^{\pm}}$,
$$k^+=k_G~,\quad k^-=\tilde{h}_G-2~, \eqno(3.16)$$
and the $N=4~$ QSCA central charge,
$$c=\fracmm{6(k_G+1)(\tilde{h}_G-1)}{k_G+\tilde{h}_G}-3~, \eqno(3.17)$$
are also fixed. All the generators and the parameters of the non-linear
algebra are now determined, and it is straightfowrard (although quite tedious)
to verify the rest of the $\hat{D}(2,1;\a)~$ QSCA. No additional consistency
conditions arise.

As far as the {\it symmetric} quaternionic spaces are concerned, eqs.~(3.4),
(3.14), (3.15) and the defining OPE's of the $\hat{D}(2,1;\a)$ algebra in
Appendix lead to very simple expressions for the generators of this non-linear
 algebra on such spaces,
$$\eqalign{
G^0=\fracmm{i}{\sqrt{2k}}\j^{\bar{a}}\hat{J}^{\bar{a}}~,\quad ~&~ \quad
G^M=\fracmm{i}{\sqrt{2k}}h^M_{\bar{a}\bar{b}}\j^{\bar{a}}\hat{J}^{\bar{b}}~,
\cr
J^{M-}=\fracmm{1}{16}h^M_{\bar{a}\bar{b}}\j^{\bar{a}}\j^{\bar{b}}~,
 \quad ~&~ \quad
J^{M+}=\fracmm{i}{4(\tilde{h}_G-2)}h^M_{\bar{a}\bar{c}}f^{\bar{a}\bar{c}d}
\hat{J}^d~,\cr}\eqno(3.18)$$
$$T=\fracmm{1}{k}\left[ \hat{J}^{\bar{a}}\hat{J}^{\bar{a}} + \ha (k_G+1)
(\j^{\bar{a}}\pa\j^{\bar{a}})-if^{i\bar{b}\bar{c}}\hat{J}^i
\j^{\bar{b}}\j^{\bar{c}} +\ha d_{\bar{a}\bar{b}\bar{c}\bar{d}}
\j^{\bar{a}}\j^{\bar{b}}\j^{\bar{c}}\j^{\bar{d}}\right]~,$$
where $d_{\bar{a}\bar{b}\bar{c}\bar{d}}$ are certain linear combinations of
the structure constants --- see the l.h.s. of eq.~(4.17) below.

Given a simple Lie group $G$, there is the {\it unique} (associated with this
group) quaternionic symmetric space, which is called the {\it Wolf} space.  To
introduce this space, let $(E_{\j\pm},H_{\j})$ be the generators of the
$su(2)_{\j}$ subalgebra of $\cg$, associated with the highest root $\j$,
$$ \[ E_{\j+},E_{\j-} \]=2H_{\j}~,\quad \[ H_{\j},E_{\j\pm} \]=\pm
E_{\j\pm}~.\eqno(3.19)$$
The associated Wolf space is the coset
$$ \fracmm{G}{H_{\bot}\otimes SU(2)_{\j}}~,\eqno(3.20)$$
where $H_{\bot}$ is a {\it centralizer} of $SU(2)_{\j}$ in $G$. The cosets
(3.20) for various groups $G$ are of dimension $4(\tilde{h}_G-2)$, and they
are all classified \cite{wolf,al}. The {\it non-symmetric} spaces
$(G/H_{\bot})\otimes U(1)$ of dimension $4(\tilde{h}_G-1)$ are also
quaternionic. Therefore, the both different sets of cosets,
$$\fracmm{G}{H_{\bot}\otimes SU(2)_{\j}}~,\qquad {\rm and} \qquad
\fracmm{G\otimes U(1)}{H_{\bot}}~,\eqno(3.21)$$
support the non-linear $N=4~$ QSCA, but only the second one supports the
 `large' linear $N=4~$ SCA too \cite{st}. The list of compact Wolf spaces and
the QSCA central charges of the associated $N=4~$ SCFT's are collected in
Table 1. The only known {\it non-compact} quaternionic spaces are just
non-compact analogues of those listed in Table 1, as well as some additional
non-symmetric spaces found by Alekseevskii \cite{al}.

\newpage

{\sf Table 1}. The Wolf spaces, and the (Virasoro) central charges of the
associated $N=4~$ SCFT's\/, with respect to the $N=4~$ $~\hat{D}(2,1;\a)~$
QSCA. Here $k^+=k_G\equiv\hat{k}$, $k^-=\tilde{h}_G-2$, and
$c_{\rm GS}=6(\hat{k}+1)(\tilde{h}_G-1)/(\hat{k}+\tilde{h}_G)-3$.
\vglue.1in
\begin{tabular}{lllllll} \hline

$G/[H_{\bot}\otimes SU(2)]$ & & dim & &$\tilde{h}_G$ &&
$c_{\rm GS}$ \\
\hline
$\fracmm{Sp(n)}{Sp(n-1)\otimes Sp(1)}$ & $n>1$ & $4n-4$ && $n+1$ &&
$6n-3-6n^2/(\hat{k}+n+1)$ \\
$\fracmm{SU(n)}{SU(n-2)\otimes SU(2)\otimes U(1)}$ & $n>2$ & $4n-8$ && $n$ &&
$3(2n-3)-6(n-1)^2/(\hat{k}+n) $ \\
$\fracmm{SO(n)}{SO(n-4)\otimes SO(4)}$ & $n>4$ & $4n-16$ && $n-2$ && $3(2n-7)
-6(n-3)^2/(\hat{k}+n-2) $   \\
$\fracmm{G_2}{SO(4)}$ && $8$ && $4$ && $9-36/(\hat{k}+4) $  \\
$\fracmm{F_4}{Sp(3)\otimes Sp(1)}$ && $28$ && $9$  && $45-384/(\hat{k}+9) $  \\
$\fracmm{E_6}{SU(6)\otimes SU(2)}$ && $40$ && $12$ && $63-726/(\hat{k}+12)
  $    \\
$\fracmm{E_7}{SO(12)\otimes SU(2)}$ && $64$  && $18$ &&
$99-1734/(\hat{k}+18)  $  \\
$\fracmm{E_8}{E_7\otimes SU(2)}$ && $112$  && $30$ && $177-5220/(\hat{k}+30)
$  \\
\hline
\end{tabular}

\vglue.2in

The `{\it small\/}' linear $N=4~$ SCA can be formally obtained from the
`large' linear $N=4~$ SCA in the limit $k^-\to\infty$ and $k^+\to 0$. We are
however not in a position to get SCFT's based on the `small' linear
$N=4~$ SCA from our $N=4$ coset construction since $k^+$ is the {\it only}
parameter at our disposal according to eq.~(3.16), which is not enough.
This simple observation already makes a difference between the `old' $N=4$
strings  \cite{ade}, based on the `small' linear $N=4~$ SCA, and the `new'
$N=4$ strings based on the non-linear $N=4~$ $~\hat{D}(2,1;\a)~$  QSCA
\cite{k4}.

The unitary highest-weight (positive energy) representations of the non-linear
algebra were investigated by G\"unaydin, Petersen, Taormina and van Proeyen
\cite{gptv}. They showed that the central charge values leading to the {\it
rational} $N=4$ SCFT's (with {\it finite} numbers of different unitary
representations) arise when $k^-=0$, for the so-called {\it massless}
representations labeled by the integer $k_G$ and the half-integral
highest-weight of the $su(2)$ subalgebra \cite{gptv}. This implies
$\tilde{h}_G=2$ in the coset approach above. According to Table 1, no such
unitary (massless) rational $N=4$ SCFT's can appear in our construction.

\newpage

\section{$N=4$ ~NLSM and WZNW}

In the previous section, we constructed the $N=4$ coset models by using the
techniques of 2d CFT. A natural question arises whether our models can be
identified with certain 2d {\it non-linear sigma-models} (NLSM's). The CFT
construction applies to the holomorphic sector of a 2d field theory which
corresponds to its left-moving degrees of freedom after the (inverse) Wick
rotation. Therefore, by $N=4$ supersymmetry above we actually mean $(4,0)$
supersymmetry.~\footnote{In two dimensions, a supersymmetry algebra can have
$p$ left-moving and $q$ right-moving \newline ${~~~~~}$ real supercharges.}
In this section, we
want to compare the $N=4$ SCFT construction with the standard two-dimensional
$N=4~$ NLSM construction known in the literature (see ref.~\cite{pt} for a
recent review), and build the relevant $N=4~$ WZNW actions on Wolf spaces.
\vglue.2in

\subsection{$(4,0)$ NLSM from the viewpoint of $(1,0)$ superspace}

Since an arbitrary bosonic NLSM can be made supersymmetric with respect to
$N=~1$ or $(1,0)$ supersymmetry, it seems to be quite natural to require an
{\it explicit} $(1,0)$ supersymmetry of the $(4,0)$ supersymmetric NLSM in
question. By `explicit' we mean `off-shell', in order to use superspace. It
should be noticed however that only {\it on-shell} supersymmetry is required
in SCFT.
Since our $N=4$ supersymmetry is going to be non-linearly realized in general,
the standard (or harmonic) $N=4$ superspace cannot be applied, at least
naively, because it implies a {\it linearly} realized $N=4$ supersymmetry,
which is too restrictive for our purposes, as we already know from the
previous section. To make contact with the standard results, we start from the
$N=1$ or $(1,0)$ supersymmetric 2d NLSM.

The $(1,0)$ superspace action for the most general $(1,0)$ NLSM reads \cite{pt}
\footnote{Our notation in this subsection is mostly self-explained, and it is
{\it different} from the one used in the \newline ${~~~~~}$
bulk of the paper.}
$$ I =\int d^2z\,d\q^+\,\left\{ (h_{ij}+b_{ij}) D_+\F^i\pa_=\F^j
+ih_{ab}\J_-^a\de_+\J^b_-\right\}~,\eqno(4.1)$$
in terms of the $(1,0)$ scalar superfields $\F^i(z^{\ip\, ,=},\q^+)$ taking
their values in a $D$-dimensional target manifold $\cm$, and the $(1,0)$ spinor
superfields $\J^a_-(z^{\ip\, ,=},\q^+)$ in a vector bundle $\ck$ over $\cm$. In
eq.~(4.1), $D_+=\fracmm{\pa}{\pa\q^+} +i\q^+\pa_{\ip}$ denotes the flat $(1,0)$
superspace covariant derivative, $\de_+\J^a_-=D_+\J^a_-+D_+\F^i\O\dud{i}{a}{b}
\J^b_-$ is its NLSM covariant generalization for the spinor superfields,
$h_{ij}(\F)$ is a metric on $\cm$, $b_{ij}(\F)$ is an antisymmetric tensor on
$\cm$, $h_{ab}(\F)$ and $\O\dud{i}{a}{b}(\F)$ are a metric and a connection
on the fibre $\ck$, respectively. It is therefore assumed that $\cm$ must be a
Riemannian manifold. In components, the action (4.2) takes the form
$$\eqalign{
 I = \int d^2z\,& \left\{  (h_{ij}+b_{ij})\pa_{\ip}\F^i\pa_=\F^j + ih_{ij}
\L^i_+\left( \pa_=\L^j_+ + \G^j_{kl}\pa_=\F^k\L^l_+ \right) \right. \cr
 & \left. -ih_{ab}\J^a_-\left( \pa_{\ip}\J^b_- + \pa_{\ip}\F^i\O\dud{i}{b}{c}
\J^c_- \right) -\ha F_{ijab}\J^a_-\J^b_-\L^i_+\L_+^j+h_{ab}F^aF^b \right\},\cr
 } \eqno(4.2)$$
where
$$\left. \F^i=\F^i\right|~,\quad \left. \L^i_+=D_+\F^i\right|~,\quad
\left. \J^a_-=\J^a_-\right|~,\quad \left. F^a=\de_+\J^a_-\right|~,\eqno(4.3)$$
and $\left.\right|$ denotes the leading component of a superfield. In eq.~(4.2)
the target space connection,
$$\G^i_{jk}=\left\{ \begin{array}{c} i \\ jk \end{array} \right\} +B^i_{jk}~,
\qquad B_{ijk}=\fracmm{3}{2}\pa_{[i}b_{jk]}~,\eqno(4.4)$$
and the fibre-valued curvature,
$$F\dud{ij}{a}{b}=\pa_i\O\dud{j}{a}{b} -\pa_j\O\dud{i}{a}{b}
+ \O\dud{i}{a}{c}\O\dud{j}{c}{b} -\O\dud{j}{a}{c}\O\dud{i}{c}{b}~,\eqno(4.5)$$
have been introduced. The scalars $F^a$ are auxiliary, and they vanish
on-shell.

The NLSM of eq.~(4.1) has manifest off-shell $(1,0)$ supersymmetry. Requiring
further (non-manifest) supersymmetries implies certain restrictions on the
NLSM couplings \cite{ghr}. The form of additional supersymmetries is fixed by
dimensional analysis:
$$\eqalign{
\d_{\ve}\F^i= ~&  i\ve^{(M)}_- h\ud{(M)i}{j}(\F)D_+\F^j~,\cr
\d_{\ve}\J^a_- = ~&  i\ve^{(M)}_- h\ud{(M)a}{b}(\F)\de_+\J^b_-~,\cr}
\eqno(4.6)$$
where some tensors $h\ud{(M)i}{j}(\F)$ and $h\ud{(M)a}{b}(\F)$ have been
introduced, and $M=1,2,3$ ({\it cf~} eq.~(3.4)). It should be noticed that the
second line of eq.~(4.6) is irrelevant on-shell where $\de_+\J^b_-=0$. The
 `canonical' $(1,0)$ supersymmetry can also be represented in the form (4.6)
with $h\ud{(0)i}{j}=\d\ud{i}{j}$ and $h\ud{(0)a}{b}=\d\ud{a}{b}$, which again,
as in the previous section, invites us to switch to the four-dimensional
notation $\m=(0,M)$.

Requiring the on-shell closure of the supersymmetry transformations (4.6)
on the scalar superfields $\F^i$ alone results in {\it the same} conditions
(3.6) and (3.13) appeared in the previous section, namely, (i) the existence
of three independent complex structures satisfying the quaternionic algebra,
and (ii) the vanishing Nijenhuis tensor!  The on-shell closure on the spinor
superfields $\J^a_-$ yields
$$F\dud{ij}{a}{b}h\ud{\m i}{[k}h\ud{\n j}{l]}=\d^{\m\n}F\dud{kl}{a}{b}~,
\eqno(4.7)$$
in addition. Generally speaking, the conditions above are {\it not} enough to
ensure the invariance of the action (4.1) with respect to the transformations
(4.6), so that it could make a difference with the CFT approach. As is well
known \cite{ghr}, the action (4.1) is actually invariant provided that, in
addition, all the complex structures are hermitian {\it and} covariantly
constant with respect to the connection (4.4),
$$\de_ih^{\m}=0~.\eqno(4.8)$$
Therefore, the most general $N=4$ supersymmetry conditions for the 2d~ NLSM's
 and the SCFT's defined on cosets are exactly the same! In geometrical terms,
the $(2,0)$ supersymmetry of the NLSM requires the holonomy of the connection
(4.4) to be a subgroup of $U(D/2)$, and the vector bundle $\ck$ to be
holomorphic \cite{ghr,pt}. The $(4,0)$ supersymmetry requires the holonomy to
be a subgroup of $Sp(D/4)\otimes Sp(1)$, and the bundle $\ck$ to be holomorphic
with respect to each complex structure. The latter is known to lead to
hyper-k\"ahlerian $(b=0)$ or quaternionic $(b\neq 0)$ manifolds, whose
dimension is always a multiple of four. The holonomy conditions just mentioned
 easily follow from
the vanishing commutator of the derivatives $\de_i$ on the complex structures
$h^{\m}$, because of eq.~(4.8).
\vglue.2in

\subsection{An $N=4$ gauged WZNW action for a Wolf space}

The NLSM construction in the previous subsection is not explicit enough to
accomodate the group-theoretical structure of the (S)CFT coset models. It is
the gauged (super) WZNW actions that actually represent the relevant 2d field
theories \cite{schn}. In ref.~\cite{gu}, G\"unaydin constructed the gauged
$N=4~$ supersymmetric WZNW theories invariant under the `large' linear $N=4$
SCA.
These gauged super WZNW theories are defined over $G\otimes U(1)$, and have
the gauged subgroup $H$ such that $G/ \left[ H\otimes SU(2)\right]$ is a Wolf
space \cite{gu}. In this subsection, we modify the construction of
ref.~\cite{gu} to get the gauged super WZNW theories over the Wolf spaces.
They are going to be invariant under the non-linear $N=4~$ QSCA
$\hat{D}(2,1;\a)$.

The standard WZNW action at level $k$ is given by $kI(g)$, where
$$ I(g) = -\fracmm{1}{4\p}\int_{\S} d^2z\,\tr\left(g^{-1}\pa g\,g^{-1}\bar{\pa}
g\right) - \fracmm{1}{12\p}\int_{B} d^3y\,\ve^{\a\b\g}\tr\left(
g^{-1}\pa_{\a}g\,g^{-1}\pa_{\b}g\,g^{-1}\pa_{\g}g\right)~,\eqno(4.9)$$
where $\pa B=\S$, $\pa=\pa_z$, $\bar{\pa}=\pa_{\bar{z}}$, and the field
$g(z,\bar{z})$ takes values in the group $G$.

The gauged WZNW action reads
$$I(g,A)=I(g) + \fracmm{1}{2\p}\int_{\S} d^2z\,\tr\left(A_z\bar{\pa}g\,g^{-1}
-A_{\bar{z}}g^{-1}\pa g + A_zgA_{\bar{z}}g^{-1} -A_zA_{\bar{z}}\right)~,
\eqno(4.10)$$
where the gauge fields $(A_z,A_{\bar{z}})$, taking their values in the Lie
algebra $\ch$ of a diagonal subgroup $H$ of the global $G_{\rm L}\otimes
G_{\rm R}$ symmetry of the WZNW action (4.9), have been introduced.

The gauged $(1,0)$ supersymmetric WZNW action for a coset $G/H$ takes the form
 \cite{schn,gkks}
$$I(g,A,\J)= I(g,A) + \fracmm{i}{4\p}\int_{\S} d^2z\,\tr\left(
\J\Bar{D}\J\right)~,\eqno(4.11)$$
where the 2d {\it Majorana-Weyl} (MW) fermions $\J^{\bar{a}}$ valued in the
 orthogonal complement $\cn$ of the Lie algebra $\ch$ in the Lie algebra $\cg$
 have
been introduced,  $\Bar{D}\J^{\bar{a}}=\bar{\pa}\J^{\bar{a}}
 + f^{\bar{a}\bar{b}d}\J^{\bar{b}}A^d_{\bar{z}}$. Compared to the most general
$(1,0)$ NLSM in eq.~(4.2), the $(1,0)$ WZNW action (4.11) does not contain
$(1,0)$ spinor multiplets and has no quartic fermionic couplings.

The gauge transformations of the fields are
$$\d g =\[u,g\]~,\quad \d A_z=Du~,\quad \d A_{\bar{z}}=\Bar{D}u~,\quad
\d\J=\[u,\J \]~,\eqno(4.12)$$
where $Du=\pa u -\[A_z,u\]$, $\Bar{D}u=\bar{\pa}u -\[A_{\bar{z}},u\]$, and
$u$ is the $\ch$-valued infinitesimal gauge parameter. The on-shell $(1,0)$
supersymmetry of the action (4.11) is
$$\d g=i\ve g\J~,\quad \d\J=\ve\left(g^{-1}D_zg -i\J^2\right)_{\cn}~,
\quad \d A=0~.\eqno(4.13)$$

The action (4.11) is a good starting point to examine further supersymmetries.
In particular, as was shown by Witten \cite{w}, that action admits $(2,0)$
supersymmetry when the coset space is {\it k\"ahlerian}, the canonical example
being provided by the grassmannian manifolds
$SU(n+m)/ \left[ SU(m)\otimes SU(n)\otimes U(1)\right]$ \cite{nak}. A
quantization of the action for k\"ahlerian cosets results in a subclass of
the KS models (subsect.~2.4), namely, those of them which have
${\rm rank}\,G={\rm rank}\,H$. According to our discussion in subsect.~2.4,
the rest of non-k\"ahlerian but still $N=2$ supersymmetric KS models
corresponds to the cases when $G/H=K\otimes U(1)^{2n}$, $n=1,2,\ldots\,$,
where
$K$ is a k\"ahlerian coset. It is trivial to generalize Witten's construction
of the $N=2$ gauged WZNW actions to the other (non-k\"ahlerian) cases, since
the factor $U(1)^{2n}$ is abelian and, therefore, it merely contributes a free
supersymmetric action for $n$ scalar $(2,0)$ supermultiplets. Without loss of
generality, we can restrict ourselves to the case of $n=0$ in our construction
of the $N=4$ actions, modulo adding a free action for some number of chiral
scalar $(4,0)$ supermultiplets.~\footnote{Free chiral scalar $N=4$
supermultiplets are still relevant in $N=4$ string theory, since they
\newline ${~~~~~}$ contribute to the conformal anomaly. They play the role
similar to free scalars \newline ${~~~~~}$ appearing in the toroidally
compactified (four-dimensional) superstrings.}

To this end, we are going to elaborate the structure of the gauged super-WZNW
theories on the Wolf spaces (3.20), by using G\"unaydin's results about coset
realizations of the $N=4$ extended SCA's over the so-called {\it Freudenthal
 triple systems} (FTS's) \cite{gu}. A convenient (Kantor) decomposition of the
Lie algebra $\cg$ is given by its decomposition into the eigenspaces with
respect to the grading operator $H_{\j}$ \cite{kan},
$$\cg = \cg^{(-2)}\oplus\cg^{(-1)}\oplus\cg^{(0)}\oplus\cg^{(+1)}\oplus
\cg^{(+2)}~,\eqno(4.14)$$
where the $H_{\j}$-eigenvalues appear as superscripts (in brackets).~\footnote{
 The elements of $\cg^{(-1)}$ can be put in one-to-one correspondence with FTS,
the latter \newline ${~~~~~}$ being usually represented by a division algebra
 \cite{freu}.} The
one-dimensional spaces $\cg^{(-2)}$ and $\cg^{(+2)}$ just comprise $E_{\j-}$
and $E_{\j+}$, respectively, whereas $\cg^{(0)}$ can be identified with
$\ch_{\bot}\oplus H_{\j}$, where $\ch_{\bot}$ is the Lie algebra of $H_{\bot}$.
Let $E_{\bar{a}\pm}$ be the generators of $\cg^{(\pm 1)}$, and
$H^{\bot}_{\bar{a}\bar{c}}$ the generators of $\ch_{\bot}$ in the
Cartan-Weyl-type basis. The non-trivial commutation relations of $\cg$
are then given by (the signs are correlated!)
$$\eqalign{
\[ E_{\bar{a}\pm},E_{\bar{c}\pm}\]=\O^{\pm}_{\bar{a}\bar{c}}E_{\j\pm}~,\quad &
 \quad
\[ E_{\bar{a}+},E_{\bar{c}-}\]=H^{\bot}_{\bar{a}\bar{c}}+\d_{\bar{a}\bar{c}}
H_{\j}\equiv H_{\bar{a}\bar{c}} ~,\cr
\[ E_{\j\pm},E_{\bar{a}\mp}\]=\O^{\mp}_{\bar{a}\bar{c}}E_{\bar{c}\pm}~,\quad &
\quad
\[ H_{\bar{a}\bar{b}},E_{\bar{c}\pm}\]=\pm f_{\bar{a}\bar{b}\bar{c}\bar{g}}
E_{\bar{g}\pm} ~,\cr
\[ H_{\bar{a}\bar{b}},H_{\bar{c}\bar{d}}\]~~=~~&~ f_{\bar{a}\bar{b}\bar{g}
\bar{c}}H_{\bar{g}\bar{d}} - f_{\bar{a}\bar{b}\bar{d}\bar{g}}
H_{\bar{c}\bar{g}}~.\cr}\eqno(4.15)$$
Here $f_{\bar{a}\bar{b}\bar{c}\bar{d}}$ are the structure constants of
$\ch_{\bot}\oplus H_{\j}$, whose (Cartan-Weyl) normalization is fixed by the
conditions \cite{gu}
$$f_{\bar{a}\bar{a}\bar{c}\bar{d}}=\left(\tilde{h}_G-2\right)
\d_{\bar{c}\bar{d}}~,\quad
f_{\bar{a}\bar{b}\bar{b}\bar{c}}=\left(\tilde{h}_G-1\right)
\d_{\bar{a}\bar{c}}~,\eqno(4.16)$$
and which satisfy the identity
$$ f_{\bar{a}\bar{b}\bar{c}\bar{d}}-f_{\bar{a}\bar{c}\bar{b}\bar{d}}
=\O^+_{\bar{a}\bar{d}}\O^-_{\bar{b}\bar{c}}~.\eqno(4.17)$$

The matrix $\O^{\pm}_{\bar{a}\bar{c}}$ introduced in eq.~(4.15) represents a
natural {\it symplectic} structure associated with a Wolf space \cite{freu},
$$(\O^{\pm})^{-1}=\O^{\mp}~,\qquad (\O^{\pm})^{\rm T}=-\O^{\pm}~.\eqno(4.18)$$
Under hermitian conjugation, one has $(E_{\j\pm})^{\dg}=E_{\j\mp}$,
$(E_{\bar{a}\pm})^{\dg}=E_{\bar{a}\mp}$ and $(\O^{\pm})^{\dg}=-\O^{\mp}$.

It is straightforward to write down the defining OPE's of the affine Lie
algebra $\hat{\cg}$, in the form adapted to the commutation relations (4.15),
namely
$$\eqalign{
E_{\bar{a}\pm}(z)E_{\bar{c}\pm}(w)~\sim~\fracmm{\O^{\pm}_{\bar{a}\bar{c}}
E_{\j\pm}(w)}{z-w}~,\quad ~&~\quad
E_{\bar{a}+}(z)E_{\bar{c}-}(w)~\sim~\fracmm{k_G\d_{\bar{a}\bar{c}}}{(z-w)^2}
+\fracmm{H_{\bar{a}\bar{c}}(w)}{z-w}~,\cr
E_{\j\pm}(z)E_{\bar{a}\mp}(w)~\sim~\fracmm{\O^{\mp}_{\bar{a}\bar{c}}
E_{\bar{c}\pm}(w)}{z-w}~,\quad~&~\quad
H_{\bar{a}\bar{b}}(z)E_{\bar{c}\pm}(w)~\sim~\pm\fracmm{
f_{\bar{a}\bar{b}\bar{c}\bar{g}}E_{\bar{g}\pm}(w)}{z-w}~,\cr}$$
$$H_{\bar{a}\bar{b}}(z)H_{\bar{c}\bar{d}}(w)~\sim~\fracmm{k_Gf_{\bar{a}\bar{b}
\bar{c}\bar{d}}}{(z-w)^2}+\fracmm{1}{z-w}\left[ f_{\bar{a}\bar{b}\bar{g}
\bar{c}}H_{\bar{g}\bar{d}}(w) - f_{\bar{a}\bar{b}\bar{d}\bar{g}}
H_{\bar{c}\bar{g}}(w)\right]~,$$
$$E_{\j+}(z)E_{\j-}(w)~\sim~\fracmm{k_G}{(z-w)^2} + \fracmm{2H_{\j}(w)}{z-w}~,
\qquad
H_{\j}(z)E_{\j\pm}(w)~\sim~\fracmm{\pm E_{\j}(w)}{z-w}~.\eqno(4.19)$$

The gauged $(4,0)$ supersymmetric WZNW action on a Wolf space (3.20) is given
by eq.~(4.11), where the gauged group $H$ has to be $H_{\bot}\otimes
SU(2)_{\j}$, and free MW fermions $\J$ should be assigned only for the FTS
generators of $\cg^{(-1)}\oplus\cg^{(+1)}$, i.e. for  $E_{\bar{a}\pm}$. The
corresponding on-shell (holomorphic) fermions, $\j^{\bar{a}\pm}(z)$,  satisfy
the canonical OPE
$$\j^{\bar{a}+}(z)\j^{\bar{b}-}(w)~\sim~-\fracmm{\d^{\bar{a}\bar{b}}}{z-w}~.
\eqno(4.20)$$

The generators of the non-linear $~\hat{D}(2,1;\a)~$ QSCA in the $N=4$ gauged
WZNW theory were identified in ref.~\cite{gu}. Compared with eq.~(3.18) in the
SCFT approach, the $N=4$ supersymmetry generators in the field theory (WZNW)
approach naturally appear in the $(2,2)$ representation instead of the $(1,3)$
 one in eq.~(3.18), namely \cite{gu}
$$\eqalign{
G^0(z)\pm iG^1(z)~=~&~ \fracmm{2}{\sqrt{k_G+\tilde{h}_G}}\,
\j^{\bar{a}\pm}(z)E_{\bar{a}\pm}(z)~,\cr
G^2(z)\pm iG^3(z)~=~&~\fracmm{\mp 2}{\sqrt{k_G+\tilde{h}_G}}\,
\j^{\bar{a}\pm}(z)\O^{\mp}_{\bar{a}\bar{c}}E_{\bar{c}\pm}(z)~.\cr}
\eqno(4.21)$$
The generators of the first $\Hat{su(2)}$ affine subalgebra (at level $k_G$)
of the QSCA are just given by the $SU(2)_{\j}$ currents $E_{\j\pm}(z)$ and
$H_{\j}(z)$ -- see the last two lines of eq.~(4.19). The generators of the
second $\Hat{su(2)}$ affine subalgebra at level $\tilde{h}_G$ are bilinears of
 free fermions \cite{gu},
$$J_{\pm}(z)=\ha\O^{\pm}_{\bar{a}\bar{c}}\j^{\bar{a}\mp}(z)\j^{\bar{c}\mp}(z)~,
\quad J_3(z)=-\ha\j^{\bar{a}+}(z)\j^{\bar{a}-}(z)~.\eqno(4.22)$$
Finally, the QSCA stress tensor reads \cite{gu}
$$\eqalign{
T=~&~\fracmm{1}{k_G+\tilde{h}_G}\left\{ \ha\left(E_{\bar{a}+}E_{\bar{a}-}
+E_{\bar{a}-}E_{\bar{a}+}\right) + \ha\left(E_{\j+}E_{\j-}
+E_{\j-}E_{\j+}\right) + H_{\j}^2 \right. \cr
{}~&~\left. +\fracmm{k_G+1}{2}\left(\j^{\bar{a}+}\pa\j^{\bar{a}-}+
\j^{\bar{a}-}\pa\j^{\bar{a}+}\right) -H^{\bot}_{\bar{a}\bar{c}}
\j^{\bar{a}+}\j^{\bar{a}-}+\ha\O^+_{\bar{a}\bar{b}}\O^-_{\bar{c}\bar{d}}
\j^{\bar{a}-}\j^{\bar{b}-} \j^{\bar{c}+}\j^{\bar{d}+}\right\}~.}
\eqno(4.23)$$

It is instructive to compare the $N=4~$ QSCA generators obtained from the $N=4$
SCFT coset approach in eq.~(3.18), with the $N=4~$ WZNW generators given above.
 First, we immediately see that they actually coincide after identifying
$$E_{\bar{a}\pm}(z)=\fracmm{i}{\sqrt{2k}}\hat{J}_{\bar{a}\pm}(z)~,\quad
{\rm ~where~} \quad  k=k_G+\tilde{h}_G~,\eqno(4.24)$$
and using the crucial identity (4.17). Second, after identifying the
generators as above, we find the quaternionic structure
$\{h^{\m}\}$, $\m=0,1,2,3$, on a Wolf space, in terms of the symplectic
structure of the associated FTS. The first complex structure takes, of course,
the canonical form, as it should,
$$h^{(1)}_{(\bar{a}\pm)(\bar{b}\pm)}=\left( \begin{array}{cc}
-i\d_{\bar{a}\bar{b}}
 & 0 \\ 0 & +i\d_{\bar{a}\bar{b}} \end{array} \right)~.\eqno(4.25)$$
As far as the other two complex structures are concerned, we find
$$\eqalign{
h^{(2)}_{(\bar{a}\pm)(\bar{b}\pm)}=\left(
\begin{array}{cc} -\O^-_{\bar{a}\bar{b}}
 & 0 \\ 0 & +\O^+_{\bar{a}\bar{b}} \end{array} \right)~,\quad ~&~ \quad
h^{(3)}_{(\bar{a}\pm)(\bar{b}\pm)}=\left(
\begin{array}{cc}+i\O^-_{\bar{a}\bar{b}}
 & 0 \\ 0 & =i\O^+_{\bar{a}\bar{b}} \end{array} \right)~.}\eqno(4.26)$$
In particular, $h^{(1)}h^{(2)}=h^{(3)}$, as it should. The dimension of a Wolf
space, $D_{\rm W}=4(\tilde{h}_G-2)$, is clearly twice the dimension of the
corresponding FTS.

Summarizing the above-mentioned in this section, the $N=4$ field theory (WZNW)
 approach leads to {\it the same} results as the $N=4~$ SCFT approach, although
 in a more tedious way.
\vglue.2in

\section{New $~N=4~$ strings}

We are now in a position to discuss $N=4$ strings propagating on Wolf spaces.
The coset realizations of the $N=4~$ QSCA considered above give relevant
constraints on the $N=4$ string physical states in the form
$$\eqalign{
\left\{ \ha\left(E_{\bar{a}+}E_{\bar{a}-} + E_{\bar{a}-}E_{\bar{a}+}\right)
+\ha (k_G+1)\left(\j^{\bar{a}+}\pa\j^{\bar{a}-}+\j^{\bar{a}-}\pa\j^{\bar{a}+}
\right)\right. ~&~ \cr
\left. - H^{\bot}_{\bar{a}\bar{c}}\j^{\bar{a}+}\j^{\bar{a}-} +
\ha\O^+_{\bar{a}\bar{b}}\O^-_{\bar{c}\bar{d}}
\j^{\bar{a}-}\j^{\bar{b}-} \j^{\bar{c}+}\j^{\bar{d}+}\right\}\ket{\rm phys}=0~,
{}~&~\cr}$$
$$\eqalign{
\j^{\bar{a}\pm}E_{\bar{a}\pm}\ket{\rm phys}=~&~
\j^{\bar{a}\pm}\O^{\mp}_{\bar{a}\bar{c}}E_{\bar{c}\pm}\ket{\rm phys}=0~,\cr
E_{\j\pm}\ket{\rm phys}=~&~ H_{\j}\ket{\rm phys}=0~,\cr
\O^{\pm}_{\bar{a}\bar{c}}\j^{\bar{a}\mp}\j^{\bar{c}\mp}\ket{\rm phys}=~&~
\j^{\bar{a}+}\j^{\bar{a}-}~\ket{\rm phys}=0~,\cr}\eqno(5.1)$$
where eqs.~(4.21), (4.22) and (4.23) have been used. It is obvious that these
constraints are very different from the ones proposed in ref.~\cite{ade}, and,
therefore, they define a new theory of $N=4$ strings. Note, in particular, a
presence of the quartic fermionic term in the second line of eq.~(5.1).
Although the string constraints (5.1) look very complicated, the $N=4$ QSCA
they satisfy actually allows us to get information about their content from
the corresponding $N=4~$ SCFT.

The full invariant 2d action for this $N=4$ string theory is obtained
by promoting the superconformal symmetries of the $N=4$ gauged WZNW action to
the local level. As is usual in string theory, the string constraints (5.1)
are to be in one-to-one correspondence with proper on-shell $N=4$ supergravity
 fields. In our case, the new $W$-type $N=4$ supergravity seems to be needed
\cite{k4}, and its gauge fields are~\footnote{Following ref.~\cite{k4}, we
call it $\tilde{D}_4$ supergravity.}
$$ e^a_{\a}~,\qquad \c^{\m}_{\a}~,\qquad B^{I\pm}_{\a}~,\eqno(5.2)$$
where $e^a_{\a}$ is a zweibein, $\c^{\m}_{\a}$ are four 2d~ MW gravitinos, and
$B^{I\pm}_{\a}$ are six $SU(2)\otimes SU(2)$ gauge fields. The full action is
obtained by adding to the rigid $N=4$ action (4.11) the Noether coupling for
the $N=4$ supersymmetry, and minimally covariantizing the result with respect
to all the gauge fields in eq.~(5.2) \cite{k4}. No additional terms are needed
 in the action.~\footnote{Of course, as is always the case in the Noether
 procedure, the transformation laws of \newline ${~~~~~}$ the fields receive
 proper modifications.} Like in the `old' invariant $N=4$ string action
found by
Pernici and Nieuwenhuizen \cite{pn}, the rigid and local $N=4$ models have
{\it the same} geometry for the internal NLSM manifold parametrized by the
scalar fields (i.e. quaternionic), and no constraints on the $Sp(1)$ curvature
of a quaternionic manifold arise, unlike in four dimensions \cite{bw}. Instead
of concentrating on the action and the transformation laws \cite{gk}, we
proceed with the BRST quantization.

The gauge field content of the $\tilde{D}_4$ conformal 2d supergravity is
balanced by the gauge symmetries as usual, which implies no off-shell degrees
of freedom (up to moduli). In quantum theory, some of the gauge
symmetries may become anomalous and thereby some of the gauge degrees of
freedom may become physical.

The BRST ghosts appropriate for this case are:
\begin{itemize}
\item the conformal ghosts ($b,c$), an anticommuting pair of
world-sheet free fermions of conformal dimensions~($2,-1$), respectively;
\item the $N=4$ superconformal ghosts ($\b^{\m},\g^{\m}$) of conformal
dimensions~($\frac32,-\frac12$), respectively, in the fundamental (vector)
representation of $SO(4)$;
\item the $SU(2)_+\otimes SU(2)_-$ internal symmetry ghosts ($\tilde{b}^{I\pm},
\tilde{c}^{I\pm}$) of conformal dimensions~($1,0$), respectively, in the
adjoint representation of $SU(2)_{\pm}$.
\end{itemize}
The conformal ghosts
$$b(z)\ =\ \sum_{n\in{\bf Z}} b_n z^{-n-2}~,\qquad
c(z)\ =\ \sum_{n\in{\bf Z}} c_n z^{-n+1}~,\eqno(5.3)$$
have the following OPE and  anticommutation relations:
$$b(z)\ c(w)\ \sim\ \fracmm{1}{z-w}~,
\qquad \{c_m,b_n\}\ =\ \d_{m+n,0}~.\eqno(5.4)$$
The superconformal ghosts
$$\b^{\m}(z)\ =\ \sum_{r\in{\bf Z}(+1/2)}\b^{\m}_r z^{-r-3/2}~,\qquad
\g^{\m}(z)\ =\ \sum_{r\in{\bf Z}(+1/2)}\g^{\m}_r z^{-r+1/2}~,\eqno(5.5)$$
satisfy
$$\b^{\m}(z)\ \g^{\n}(w)\ \sim\ \fracmm{-\d^{\m\n}}{z-w}~, \qquad
\[\g^{\m}_r,\b^{\n}_s\]\ =\d^{\m\n} \d_{r+s,0}~.\eqno(5.6)$$
An integer or half-integer moding of these generators corresponds to the usual
distinction between the Ramond- and Neveu-Schwarz--type sectors. Finally, the
fermionic $SU(2)_{\pm}$ ghosts
$$\tilde{b}^{I\pm}(z)\ =\ \sum_{n\in{\bf Z}}\tilde{b}^{I\pm}_n z^{-n-1}~,
\qquad
\tilde{c}^{I\pm}(z)\ =\ \sum_{n\in{\bf Z}} \tilde{c}^{I\pm}_n z^{-n}~,
\eqno(5.7)$$
have
$$\tilde{b}^{I\pm}(z)\ \tilde{c}^{J\pm}(w)\ \sim\ \fracmm{\d^{IJ}}{z-w}~,
\qquad
\{\tilde{c}^{I\pm}_m,\tilde{b}^{J\pm}_n\}\ =\ \d^{IJ}\d_{m+n,0}~.\eqno(5.8)$$

The BRST charge $Q_{\rm BRST}=\oint_0 \fracmm{dz}{2\p i}\,j_{\rm BRST}(z)~$
was calculated in ref.~\cite{k4}. The BRST current $j_{\rm BRST}(z)$ takes the
 form (modulo total derivative)
$$\eqalign{
j_{\rm BRST}(z)=~&~ cT + \g^{\m}G^{\m} + \tilde{c}^{I\pm}J^{I\pm} + bc\pa c
- c\tilde{b}^{I\pm}\pa\tilde{c}^{I\pm}
-\ha c\g^{\m}\pa\b^{\m}-\fracm{3}{2}c\b^{\m}\pa\g^{\m}  -b\g^{\m}\g^{\m} \cr
&~ -\ha\tilde{c}^{I\pm}(t^{I\pm})^{\m\n}\b^{\m}\g^{\n}
+ \left[ \tilde{b}^{I+}(t^{I+})^{\m\n}
  +  \tilde{b}^{I-}(t^{I-})^{\m\n}\right](\g^{\m}\pa\g^{\n}-\g^{\n}\pa\g^{\m})
 \cr
& -\ha \ve^{IJK}\tilde{c}^{I+}\tilde{c}^{J+}\tilde{b}^{K+}
 -\ha \ve^{IJK}\tilde{c}^{I-}\tilde{c}^{J-}\tilde{b}^{K-}
 +\fracmm{1}{2}\L^{\m\n}_{(I\pm)(J\pm)}J^{(I\pm)}\tilde{b}^{(J\pm)}\g^{\m}
\g^{\n} \cr
&   -\fracmm{1}{24}\L^{\m\n}_{(I\pm)(J\pm)}\L^{\l\r}_{K(\pm)L(\pm)}
 \ve^{IKN}\tilde{b}^{J(\pm)}\tilde{b}^{L(\pm)}(\tilde{b}^{N+}+\tilde{b}^{N-})
\g^{\m}\g^{\n}\g^{\l}\g^{\r}~,\cr}\eqno(5.9)$$
where the constant `non-linearity' tensor $\L$ can be easily read off  from
the last term on the r.h.s. of the supersymmetry algebra (A.1c) after rewriting
 it in terms of the self-dual currents defined by eq.~(3.2).

The quantum BRST charge (5.9) is nilpotent if and only if \cite{k4}
$$k^+=k^-=-2~,\eqno(5.10)$$
which implies, in particular
$$c_{\rm tot}\equiv c_{\rm matter}+c_{\rm gh}
=\left[\fracmm{6(k^++1)(k^-+1)}{k^++k^-+2}-3\right]+6=0~.\eqno(5.11)$$
In calculating the ghost contributions to the central charge, we used the
standard formula of conformal field theory \cite{book},
$$\eqalign{
c_{\rm gh}& = 2\sum_{\l} n_{\l}(-1)^{2\l+1}\left(6\l^2-6\l+1\right)\cr
& = 1\times (-26) + 4\times (+11) + \ha 4(4-1)\times (-2) = +6~,
\cr}\eqno(5.12)$$
where $\l$ is the conformal dimension and $n_{\l}$ is the number of the
conjugated ghost pairs: $\l=2,3/2,1$ and $n_{\l}=1,4,6$, respectively.

To cancel the positive ghost contribution, we need therefore the negative
central charge $(-6)$ for a matter representation. According to Table I, the
level $k_G$ is also negative for a negative central charge. This simple
observation already excludes unitary representations of the $N=4~$ QSCA, and,
hence, the physical space defined by
the  constraints (5.1) has little chance to be positive definite. Moreover,
comparing eqs.~(3.16) and (5.10) in the case of a Wolf space to be used as the
background for the $N=4$ string propagation, we conclude that $\tilde{h}_G=0$.
Therefore, the group $G$ has to be abelian. It leaves us only (locally flat)
tori as the consistent $N=4$ string backgrounds.
\vglue.2in

\section{Conclusion and Outlook}

Our main results are given by the title and the abstract. Contrary to the
conventional approach to $N=4$ strings based on the `small\/' $N=4~$ SCA, we
used the non-linear $N=4$ supersymmetric QSCA, which is more general. We
generalized the supersymmetric coset construction to that $N=4$ case,
constructed the relevant $N=4$ gauged WZNW actions, and defined the BRST
quantized theory of $N=4$ strings propagating on the Wolf spaces.
 Due to the non-linearity of the underlying gauged algebra, it is not possible
to build new representations by `tensoring' the known ones, similarly to
representations of $W$ algebras. Still, even that rather general framework
didn't save us from the disaster: the Wolf spaces as the $N=4$ string
backgrounds are forbidden by the quantum BRST charge nilpotency conditions, as
we showed. The only spaces allowed are just tori, which are locally flat. The
 result is rather surprising since the Wolf spaces naturally appear as
solutions in the $N=4$ coset construction. Consistent backgrounds for the $N=4$
 string propagation may also exist outside cosets.

To this end, we would like to comment on the issue of {\it off-shell}
 extensions of the $N=4$ gauged WZNW actions. All our considerations above
were merely {\it on-shell}, which was important in our general analysis. In
particular, the super WZNW theories on the Wolf spaces are only invariant
under the on-shell $N=4$ supersymmetry which is given by the on-shell current
algebra, and which is non-linearly realised. In terms of the transformation
laws for the super WZNW fields, the non-linearity implies certain field
dependence of the `structure constants' in the commutator of two $N=4$
supertransformations. In order to get an off-shell description if any, it is
the necessary first step that the $N=4$ supersymmetry should be linearized. It
 has been known for some time \cite{gs,st,vp} that it is indeed possible,
although not for the super WZNW theories on the Wolf spaces $W$, but for those
 on cosets of the type $W\otimes SU(2)\otimes U(1)$ ({\it cf} eq.~(3.21)),
where the additional fields belonging to the $SU(2)\otimes U(1)$ group factor
serve as the `auxiliaries' to linearize the on-shell current algebra. Given
 the linear $N=4$ supersymmetry, the natural way for an off-shell
approach would be to use $N=4$ superspace. However, it is {\it not} known how
to formulate the $N=4$ super WZNW theory on a non-trivial Wolf space in $N=4$
superspace, even without coupling to any 2d supergravity theory \cite{rss}.
The related problem recently discovered \cite{gr} is a variety of ways to
define an on-shell $N=4$ scalar supermultiplet, as well as its off-shell
realizations, in two dimensions. The $N=4$ superspace constraints for scalar
supermultiplets are of most importance, since they simultaneously determine
kinematics of the propagating fields. Clearly, there are still some unsolved
problems around \cite{gk}.
\vglue.2in

\noindent{\Large\bf Acknowledgements}

One of the authors (S.V.K.) wishes to acknowledge the hospitality at the
Department of Physics, University of Maryland at College Park, extended to him
diring the initial period of this work. Useful comments from Norbert Dragon and
Olaf Lechtenfeld are appreciated.

\newpage

\noindent{\Large\bf Appendix: $\hat{D}(1,2;\a)$ QSCA and `large' $N=4~$ SCA}
\vglue.2in

The non-trivial OPE's of the $\hat{D}(1,2;\a)$ QSCA are given by \cite{gs}
$$T^{\m\n}(z)G^{\r}(w)~\sim~ \fracmm{1}{z-w}\left[\d^{\m\r}G^{\n}(w)
-\d^{\n\r}G^{\m}(w)\right]~,\eqno(A.1a)$$
$$\eqalign{
J^{\m\n}(z)J^{\r\l}(w)~\sim~& \fracmm{1}{z-w}\left[\d^{\m\r}J^{\n\l}(w)-
\d^{\n\r}J^{\m\l}(w)+\d^{\n\l}J^{\m\r}(w)-\d^{\m\l}J^{\n\r}(w)\right]\cr
& - \ha(k^+ + k^-)\fracmm{\d^{\m\r}\d^{\n\l}-\d^{\m\l}\d^{\n\r}}{(z-w)^2}
 - \ha(k^+ - k^-)\fracmm{\ve^{\m\n\r\l}}{(z-w)^2}~,\cr}\eqno(A.1b)$$
$$\eqalign{
G^{\m}(z)G^{\n}(w)~\sim~~~ &~\fracmm{4k^+k^-}{(k^+ + k^-+2)}
\fracmm{\d^{\m\n}}{(z-w)^3}+\fracmm{2T(w)\d^{\m\n}}{z-w}\cr
&  - \fracmm{k^+ + k^-}{k^+ + k^- +2}
\left[\fracmm{2J^{\m\n}(w)}{(z-w)^2} +\fracmm{\pa J^{\m\n}(w)}{z-w}\right]\cr
& + \fracmm{k^+ - k^-}{k^+ + k^- +2}\ve^{\m\n\r\l}
\left[\fracmm{J^{\r\l}(w)}{(z-w)^2} +
\fracmm{\pa J^{\r\l}(w)}{2(z-w)}\right] \cr
& - \fracmm{\ve^{\m\r\l\z}\ve^{\n\r\t\o}}{2(k^+ + k^- +2)}
\fracmm{:J^{\l\z}J^{\t\o}:(w)}{(z-w)}~.\cr} \eqno(A.1c)$$
The antisymmetric tensor $J^{\m\n}(z)$ in the adjoint of $SO(4)$ can be
decomposed into its self-dual $SU(2)$ components, see eqs.~(3.2) and (3.3).

The OPE's describing the action of $J^{M\pm}(z)$ read
$$\eqalign{
J^{M\pm}(z)J^{N\pm}(w)~\sim~ & \fracmm{\ve^{MNP}J^{P\pm}(w)}{z-w} +
\fracmm{-k^{\pm}\d^{MN}}{2(z-w)^2}~,\cr
J^{M\pm}(z)G^{\m}(w)~\sim~ &
\fracmm{{\frac 12}(t^{M\pm})^{\m\n}G^{\n}(w)}{z-w}~,\cr}\eqno(A.2)$$
where {\it two} arbitrary `levels' $k^{\pm}$ for both independent
$\Hat{su(2)}_{\pm}$ affine Lie algebra components have been introduced.

Though $\hat{D}(1,2;\a)$ is a non-linear QSCA, it can be turned into a
{\it linear} SCA by adding some `auxiliary' fields, namely, four free fermions
 $\j^{\m}(z)$ of dimension $1/2$, and a free bosonic current $U(z)$ of
dimension $1$ \cite{gs}. The new fields have canonical OPE's,
$$\j^{\m}(z)\j^{\n}(w)~\sim~ \fracmm{-\d^{\m\n}}{z-w}~,\qquad
U(z)U(w)~\sim~ \fracmm{-1}{(z-w)^2}~.\eqno(A.3)$$
The fermionic fields $\j^{\m}(z)$ transform in $(2,2)$ representation of
$SU(2)_+\otimes SU(2)_-$,
$$J^{M\pm}(z)\j^{\m}(w)~\sim~ \fracmm{{\frac 12}(t^{M\pm})^{\m\n}\j^{\n}(w)}{z
-w}~,\eqno(A.4)$$
whereas the singlet $U(1)$-current $U(z)$ can be thought of as derivative of
a free scalar boson, $U(z)=i\pa\f(z)$. The new currents takes the form
\cite{gs}
$$\eqalign{
T_{\rm tot} = & T -\ha :U^2: - \ha:\pa\j^{\m}\j^{\m}:~,\cr
G^{\m}_{\rm tot} = & G^{\m} - U\j^{\m}
+\fracmm{1}{3\sqrt{2(k^+ +k^-+2)}}\ve^{\m\n\r\l}\j^{\n}\j^{\r}\j^{\l}\cr
&  -\sqrt{\fracmm{2}{k^+ +k^-+2}}\,\j^{\n}\left[(t^{M+})^{\n\m}J^{M+}-
(t^{M-})^{\n\m}J^{M-}\right]~,\cr
J^{M\pm}_{\rm tot} = & J^{M\pm} +\fracmm{1}{4}(t^{M\pm})^{\m\n}\j^{\m}\j^{\n}~,
\cr}\eqno(A.5)$$
in terms of the initial $\hat{D}(1,2;\a)$ QSCA currents $T~,\,G^{\m}$ and
$J^{M\pm}$. It follows that the generators $T_{\rm tot}~$, $G^{\m}_{\rm tot}~$,
 $J^{M\pm}_{\rm tot}~$, $\j^{\m}~$ and $U$ have closed OPE's among themselves,
and define a `large' linear $N=4$ SCA with
$\Hat{su(2)}\oplus\Hat{su(2)}\oplus\Hat{u(1)}$ affine Lie subalgebra \cite{gs}.
The non-trivial OPE's of the `large' linear $N=4$ SCA are
$$\eqalign{
T_{\rm tot}(z)T_{\rm tot}(w)~\sim~ & \fracmm{{\frac 12}(c+3)}{(z-w)^4}
 + \fracmm{2T_{\rm tot}(w)}{(z-w)^2} + \fracmm{\pa T_{\rm tot}(w)}{z-w}~,\cr
T_{\rm tot}(z)\co(w)~\sim~ & \fracmm{h_{\co}\co(w)}{(z-w)^2} +
 \fracmm{\pa \co (w)}{z-w}~,\cr
J_{\rm tot}^{M\pm}(z)J_{\rm tot}^{N\pm}(w)~\sim~ &
\fracmm{\ve^{MNP}J_{\rm tot}^{P\pm}(w)}{z-w} - \fracmm{(k^{\pm}+1)\d^{MN}}{2
(z-w)^2}~,\cr
J_{\rm tot}^{M\pm}(z)G^{\m}_{\rm tot}(w)~\sim~ &
\fracmm{{\frac 12}(t^{M\pm})^{\m\n}G^{\n}_{\rm tot}(w)}{z-w}  \mp
\fracmm{k^{\pm}+1}{\sqrt{2(k^++k^-+2)}}
\fracmm{(t^{M\pm})^{\m\n}\j^{\n}(w)}{(z-w)^2}~,\cr
G^{\m}_{\rm tot}(z)G^{\n}_{\rm tot}(w)~\sim~ &
\fracmm{{\frac 23}(c+3)\d^{\m\n}}{(z-w)^3} +\fracmm{2T_{\rm tot}(w)\d^{\m\n}}{
z-w} -\fracmm{2}{k^++k^-+2}\left[\fracmm{2}{(z-w)^2}+\fracmm{1}{z-w}\pa_w
\right]\cr
& \times \left[(k^-+1)(t^{M+})^{\m\n}J^{M+}_{\rm tot}(w)+
(k^++1)(t^{M-})^{\m\n}J^{M-}_{\rm tot}(w)\right]~,\cr
\j^{\m}(z)G^{\n}_{\rm tot}(w) ~\sim~ & \fracmm{1}{z-w}\sqrt{
\fracmm{2}{k^++k^-+2}}\left[ (t^{M+})^{\m\n}J^{M+}_{\rm tot}(w)
-(t^{M-})^{\m\n}J^{M-}_{\rm tot}(w)\right]\cr
 ~&~ +\fracmm{U(w)\d^{\m\n}}{z-w}~,\cr
U(z)G_{\rm tot}^{\m}(w) ~\sim~ & \fracmm{\j^{\m}(w)}{(z-w)^2}~,\cr}\eqno(A.6)$$
where $\co$ stands for the generators $G_{\rm tot},\,J_{\rm tot}$ and $\j$ of
dimension $3/2,\,1$ and $1/2$. The $\hat{D}(1,2;\a)$ QSCA central charge is
 $$c=\fracmm{6(k^++1)(k^-+1)}{k^++k^-+2}-3~.\eqno(A.7)$$

We define the $\a$-parameter of the $\hat{D}(1,2;\a)$ QSCA as a ratio of its
two affine `levels', $\a\equiv  k^-/k^+~$,
which measures the relative asymmetry between the two $\Hat{su(2)}$ affine Lie
algebras. When $\a=1$, i.e. $k^-=k^+\equiv k$, the $\hat{D}(1,2;1)$ QSCA
coincides with the $SO(4)$ Bershadsky-Knihznik QSCA \cite{be,kn}. The `levels'
and the central charges of those QSCA's are different,
$ k^{\pm}_{\rm large} = k^{\pm}+1$ and $c_{\rm large} = c+3$. The exceptional
 `small' $N=4$ SCA with the $\Hat{su(2)}$ affine Lie algebra component
\cite{ade} follows from the `large' $N=4$ SCA in the limit $\a\to\infty$ or
$\a\to 0$, where either $k^-\to\infty$ or $k^+\to\infty$, respectively. Taking
 the limit results in the central charge $c_{\rm small}=6k$, where $k$ is an
arbitrary `level' of the remaining $\Hat{su(2)}$ component.

\vglue.2in

\end{document}